\documentclass[preprints,article,accept,oneauthor]{Definitions/mdpi}

\firstpage{1} 
\pubvolume{xx}
\issuenum{1}
\articlenumber{5}
\pubyear{2020}
\copyrightyear{2020}
\history{}

\usepackage{amsmath,amssymb}

\Title{Nonminimal Lorentz Violation in Macroscopic Matter}

\Author{Matthew Mewes\orcidA{}}
\AuthorNames{Matthew Mewes}
\address{%
  Physics Department,
  California Polytechnic State University,
  San Luis Obispo, California 93407, USA;
  mmewes@calpoly.edu}

\abstract{
    The effects of Lorentz and CPT violations
    on macroscopic objects are explored.
    Effective composite coefficients for Lorentz violation
    are derived in terms of coefficients for electrons, protons,
    and neutrons in the Standard-Model Extension,
    including all minimal and nonminimal violations.
    The hamiltonian and modified Newton's second law
    for a test body are derived.
    The framework is applied to free-fall and torsion-balance
    tests of the weak equivalence principle
    and to orbital motion.
    The effects on continuous media are studied,
    and the frequency shifts in acoustic resonators are calculated.
}

\def\al{\alpha}
\def\be{\beta}
\def\ga{\gamma}
\def\de{\delta}
\def\ep{\epsilon}
\def\ve{\varepsilon}
\def\ze{\zeta}
\def\et{\eta}
\def\th{\theta}

\def\rh{\rho}
\def\vr{\varrho}
\def\si{\sigma}

\def\ph{\phi}
\def\vp{\varphi}
\def\ch{\chi}
\def\ps{\psi}
\def\om{\omega}
\def\Ga{\Gamma}
\def\De{\Delta}

\def\Om{\Omega}
\def\mn{{\mu\nu}}

\def\prt{\partial}
\def\cl{{\cal L}}
\def\half{\tfrac12}
\def\vev#1{\langle {#1}\rangle}

\newcommand{\rf}[1]{(\ref{#1})}

\def\mbf#1{\boldsymbol #1}
\DeclareMathOperator{\re}{Re}
\DeclareMathOperator{\im}{Im}

\def\bc#1#2{\left(\begin{smallmatrix} #1 \\ #2 \end{smallmatrix}\right)}

\def\ahat{\hat a}
\def\chat{\hat c}

\def\K{{\mathcal K}}
\def\a#1#2{{a_w}^{(#1)}_{#2}}
\def\c#1#2{{c_w}^{(#1)}_{#2}}
\def\aw#1#2#3{{a_{#1}}^{(#2)}_{#3}}
\def\cw#1#2#3{{c_{#1}}^{(#2)}_{#3}}
\def\alab#1#2{{a_w}^{(#1)\text{lab}}_{#2}}

\def\asun#1#2{{a_w}^{(#1)\text{Sun}}_{#2}}

\def\at#1#2{{a_{npe}}^{(#1)}_{#2}}
\def\ct#1#2{{c_{npe}}^{(#1)}_{#2}}
\def\att#1#2{{a'_{npe}}^{(#1)}_{#2}}
\def\ctt#1#2{{c'_{npe}}^{(#1)}_{#2}}
\def\attlab#1#2{{a'_{npe}}^{(#1)\text{lab}}_{#2}}
\def\cttlab#1#2{{c'_{npe}}^{(#1)\text{lab}}_{#2}}

\def\nr{\text{NR}}
\def\anr#1{{a_w}^{\nr}_{#1}}
\def\cnr#1{{c_w}^{\nr}_{#1}}

\def\cT#1{{c}^T_{#1}}
\def\cTlab#1{{c}^{T,\text{lab}}_{#1}}
\def\cTorb#1{{c}^{T,\text{orb}}_{#1}}

\def\bigvev#1{\big\langle{#1}\big\rangle}

\def\E#1{E_{#1}}

\def\H{{\mathcal H}}
\def\P{{\mathcal P}}
\def\S{{\mathcal S}}

\def\eff{{\rm eff}}

\def\G#1#2#3#4{\Ga^{(#1)#3#4}_{#2}}

\def\yjm#1{Y_{#1}}
\def\syjm#1#2{{}_{#1}Y_{#2}}
\def\Yjm#1{{\mathcal Y}_{#1}}
\def\Yrjm#1#2{{\mathcal Y}^{#1}_{#2}}
\def\Yrjmconj#1#2{{\mathcal Y}^{#1*}_{#2}}

\def\A#1#2{{{\mathcal A}^{#1}_{#2}}}
\def\B#1#2{{{\mathcal B}^{#1}_{#2}}}
\def\D#1#2{{{\mathcal D}^{#1}_{#2}}}

\def\del{{\mbf\nabla}}
\def\pvec{{\mbf p}}
\def\phat{\hat\pvec}
\def\xvec{{\mbf x}}

\def\vvec{{\mbf v}}
\def\vhat{\hat\vvec}
\def\Fvec{{\mbf F}}

\def\fvec{{\mbf f}}

\def\gvec{{\mbf g}}

\def\vevec{{\mbf\ve}}

\def\bevec{{\mbf\be}}
\def\behat{{\hat\bevec}}

\def\Pvec{{\mbf\P}}
\def\Phat{\hat\Pvec}
\def\uvec{{\mbf u}}
\def\uhat{\hat\uvec}
\def\Dvec{{\mbf D}}
\def\sivec{{\mbf\si}}

\def\nvec{\mbf n}
\def\nhat{\hat\nvec}
\def\evec{{\mbf e}}
\def\ehat{{\hat\evec}}
\def\Ex{\hat\evec_x}
\def\Ey{\hat\evec_y}
\def\Ez{\hat\evec_z}
\def\Eth{\hat\evec_\th}
\def\Eph{\hat\evec_\ph}
\def\Er{\hat\evec_r}

\def\Epm{\hat\evec_\pm}

\def\ER{\hat\evec^r}

\def\EMP{\hat\evec^\mp}
\def\u{\text{\scalebox{0.8}{$\uparrow$}}}
\def\d{\text{\scalebox{0.8}{$\downarrow$}}}
\def\Eu{\hat\evec_\u}
\def\Ed{\hat\evec_\d}
\def\Erh{\hat\evec_\rh}
\def\Evp{\hat\evec_\vp}

\def\omt{{\om_\text{r}}}
\def\mt{{m_\text{r}}}
\def\ms{{m_\text{s}}}
\def\ma{{m_\text{a}}}

\begin{document}

\section{Introduction}

Lorentz invariance is one of the few principles
at the heart of both General Relativity (GR) and
the Standard Model (SM) of particle physics.
However, attempts to reconcile gravity
with quantum mechanics suggest this fundamental symmetry
of Nature may be broken slightly at low energies
\cite{ks,kp}.
While Lorentz violations are expected to be minuscule,
simple estimates imply they may be
within the reach of high-precision experiments.
Spurred by this observation
and the development of the Standard-Model Extension (SME)
\cite{ck1,ck2,smegrav},
hundreds of searches for Lorentz violations in a wide variety of systems
have been performed in recent decades
\cite{bluhm_rev,tasson_rev,hees_rev,tables}.

The SME is a framework that is designed to characterize
all realistic violations of Lorentz and CPT invariance
in an effective field theory.
It contains both the SM and GR as a Lorentz-invariant limit,
which is augmented by all possible terms
involving conventional fields.
The SME  includes terms that violate Lorentz invariance and CPT invariance
as well as other fundamental principles,
such as diffeomorphism invariance \cite{rbk05,rbk08}
and the equivalence principle \cite{qbk06}.
A term in the SME action consists of combinations of
SM fields, the spacetime metric $g_\mn$, and their derivatives
contracted with a tensor coefficient for Lorentz violation
to form an observer-independent coordinate scalar.
The coefficients for Lorentz violation may vary in space and time
and could be dynamical in nature.
This is especially important when considering
Lorentz violations in GR
\cite{smegrav,rbk05,rbk08,qbk06}.
However, empirical studies generally assume
the coefficients for Lorentz violation are constant
in inertial frames,
in which case the coefficients impart a nontrivial
tensor structure to the vacuum.
The dynamics of particles and fields are altered by
interactions with this Lorentz-violating background.

A term in the action of the SME is classified, in part,
by the mass dimension $d$ of its conventional piece. 
The restriction to the lowest dimensions $d=3$ and $d=4$
is call the minimal SME \cite{ck1,ck2,smegrav}.
The full theory contains an infinite series of terms with $d\geq 3$
\cite{km09,km12,km13,km16},
which when taken together should
encompass the low-energy effective limit of any
fundamental theory unifying gravity and particle physics.
The effects of Lorentz violation typically scale
by $d$-dependent powers of the energy and momentum.
As a result,
higher-energy particles generally
give better sensitivity to nonminimal $d>4$ violations.
Most tests involving ordinary matter are highly nonrelativistic,
reducing their sensitivity to nonminimal violations.
However,
since the energy is bounded below by the mass of the particle,
a subset of effects remain finite in the limit of zero velocity.
Nonrelativistic experiments like those discussed below
are particularly sensitive to these forms of Lorentz violation.

In this work,
we examine the effects of Lorentz violation
on macroscopic matter due
to microscopic violations in free Dirac fermions.
The primary goal is to connect
Lorentz violations of arbitrary $d$
in electrons, protons, and neutrons to
signals in large objects comprised
of these particles.

Ignoring violations that lead to spin-dependent effects,
the modified Dirac lagrangian
for a fermion of species $w$
is given by \cite{km13}
\begin{eqnarray}
\cl_w &=& 
\half \bar\ps_w
(\ga^\nu i\prt_\mu - M_w) \ps_w
- \half \bar\ps_w(\ahat_w{}^\mu_{\eff} - \chat_w{}^\mu_{\eff}) \ga_\mu \ps_w 
+ {\rm h.c.}
\label{lag}
\end{eqnarray}
The Lorentz violation is controlled by the operators
$\ahat_w{}^\mu_{\eff}$ and $\chat_w{}^\mu_{\eff}$
which depend on the four-momentum $p_\mu = i\prt_\mu$.
Expanding in $p_\mu$, they take the form
\begin{eqnarray}
\ahat_w{}_\eff^\mu
&=& \sum_d a_w{}_\eff^{(d)\mu\al_1\ldots\al_{d-3}}p_{w\al_1}\ldots p_{w\al_{d-3}} \ ,
\notag\\
\chat_w{}_\eff^\mu
&=& \sum_d c_w{}_\eff^{(d)\mu\al_1\ldots\al_{d-3}}p_{w\al_1}\ldots p_{w\al_{d-3}} \ ,
\label{hat_ops}
\end{eqnarray}
where
$a_w{}_\eff^{(d)\mu\al_1\ldots\al_{d-3}}$ and
$c_w{}_\eff^{(d)\mu\al_1\ldots\al_{d-3}}$
are constant tensor coefficients for Lorentz violation.
The $a_w{}_\eff^{(d)}$ coefficients are limited to odd $d\geq3$.
They violate CPT in addition to Lorentz invariance
and can affect particles and antiparticles differently. 
The $c_w{}_\eff^{(d)}$ coefficients are nonzero for $d=\text{even} \geq4$.
They are CPT even and generally produce
the same effects in particles and antiparticles.

The above theory yields a modified Dirac equation
for electrons, protons, and neutrons,
which affects the dynamics of any object made of these particles.
The result for ordinary matter is a modified Newton's second law,
which depends on macroscopic coefficients $\cT{}$
for the observed test body $T$.
The $\cT{}$ coefficients are linear combinations
of the $a_w{}_\eff^{(d)}$ and $c_w{}_\eff^{(d)}$
coefficients for electrons, protons, and neutrons.
These combinations depend on the relative numbers of particles of each species.
So different forms of matter with different particle content can,
in principle, be used to disentangle the violations
in different species.

The $d=3$ and $d=4$ violations in \rf{lag}
are part of the minimal SME \cite{ck1,ck2}
and have received intense scrutiny in the intervening decades
since its construction \cite{tables}.
It has been shown that the $d=3$ violations
associated with the $a_w{}_\eff^{(3)\mu}$ coefficients
can be removed from the theory through a field redefinition
and have no physical effects \cite{ck1}.
We will therefore restrict attention to violations with $d\geq 4$.
Note, however, that $a_w{}_\eff^{(3)\mu}$ violations
are observable through Lorentz-violating matter-gravity couplings \cite{kt09}.
The $d=4$ coefficients $c_w{}_\eff^{(4)\mn}$ are observable.
They do, however, mimic a species-specific defect in
the spacetime metric $\et^\mn$,
which can be removed from one particle
through a coordinate transformation \cite{km09,ctrans1,ctrans2}.
We use this freedom to eliminate analogous coefficients
from the photon sector.
Other minimal violations in photons produce birefringence
and are strictly limited by astrophysical tests
\cite{bire1,bire2,bire3,bire4,bire5,bire6,bire7,bire8,bire9,bire10,bire11}.
We therefore can safely neglect the effects of minimal Lorentz violations
in the pure-photons sector of the SME.
We will also neglect violations in
electromagnetic interactions \cite{qed1,qed2},
matter-gravity couplings \cite{kt09,kt11,li20},
and nonminimal violations in photons \cite{km09}.
Including these would be of interest
but would complicate the analysis.
They are expected to produce similar effects to those found here.

To date,
constraints on minimal
$c_w{}_\eff^{(4)}$ coefficients
have been placed in studies involving
astrophysics
\cite{astro1,astro2,astro3,astro4,astro5,km13,kt11,shao_matter,pulsar1},
tests of the equivalence principle \cite{wep1,wep2,kt11},
gravimeters \cite{gravimeters},
accelerators \cite{accel1,accel2,accel3},
electromagnetic cavities \cite{cavities1,cavities2,cavities3},
atomic systems
\cite{atoms1,atoms2,atoms3,atoms4,atoms5,atoms6,atoms7,
  atomse1,atomse2,atomse3,atomspn1,atomspn2},
and
acoustic resonators \cite{quartz1}.
The sensitivities in electrons
have reached levels of parts in $10^{20}$
in experiments involving atomic clocks \cite{atomse1}
and trapped ions \cite{atomse2,atomse3}.
Constraints on protons and neutrons have
reached the $10^{-29}$ level
in tests using comagnetometers \cite{atomspn1,atomspn2}.
Many of the atomic constraints have
been translated into similarly stringent bounds
on nonminimal $d>4$ violations \cite{kv15,kv18}.
Tight constraints on nonminimal violations
have also been inferred from
laboratory \cite{schreck16}
and astrophysical \cite{km09}
tests of relativistic kinematics
and from Penning-trap experiments \cite{ding20}.
See \cite{tables}
for an extensive list of constraints
on Lorentz violation in other sectors.
While experiments based on microscopic physics
give sensitivities that are orders of magnitude
beyond what has been demonstrated with macroscopic matter,
each experiment is based on different assumptions.
So macroscopic tests of Lorentz invariance
play an important complementary role.

This paper is organized as follows.
The basic theory is discussed in Section \ref{THEORY}.
An effective hamiltonian for ordinary macroscopic
matter is constructed in Section \ref{HAMILTONIAN},
and a modified Newton's second law is derived in Section \ref{EOM}.
Section \ref{LT} provides a brief review of observer rotations
of spherical SME coefficients
and derives, for the first time,
the boosts of the spherical coefficients
to first order in boost velocity.
Several applications are discussed in Section \ref{APPLICATIONS},
including tests of the weak equivalence principle in Section \ref{WEP}
and tests involving planetary orbits in Section \ref{ORBITS}.
A lagrangian for continuous media is given in Section \ref{RES}
and used to derive the frequency shift in
piezoelectric acoustic resonators.
Section \ref{SUMMARY} summarizes the results of the work.
A useful product identity for
spherical-harmonic tensors
is derived in the Appendix.

\section{Theory}
\label{THEORY}

\subsection{Hamiltonian}
\label{HAMILTONIAN}

Ignoring spin-dependent violations,
the leading-order effects of Lorentz violation
on a free Dirac fermion of species $w$
are described by the effective hamiltonian $h_w = \E{w} + \de h_w$,
where \cite{km13}
\begin{equation}
\de h_w = \E{w}^{-1}(\ahat_w{}^\mu_{\eff} - \chat_w{}^\mu_{\eff})p_{w\mu} \ .
\label{hw}
\end{equation}
Here, $\E{w}=\sqrt{\pvec_w^{\,2}+M_w^2}$
is the conventional free-particle energy
for species mass $M_w$.
The hamiltonian for antiparticles is given by
\rf{hw} with the opposite sign on $\ahat_w{}^\nu_{\eff}$.
Other forms of the hamiltonian \rf{hw}
may be convenient in practice.
A common signal in searches for Lorentz violation
is unexpected direction dependence,
indicating a violation of rotational symmetry.
The prominent role played by rotations in the field
motivates the spherical-harmonic expansion
\begin{equation}
\de h_w =
\sum_{dkjm}
\E{w}^{d-3-k}
|\pvec_w|^k\,
\syjm{}{jm}(\phat_w)\, (\a{d}{kjm} - \c{d}{kjm}) \ ,
\label{hr}
\end{equation}
where $0\leq k\leq d-2$, $k-j=\text{even}\geq 0$,
$|m|\leq j$, and
$\phat_w=\pvec_w/|\pvec_w|$.
The relativistic spherical coefficients for Lorentz violation
$\a{d}{kjm}$ and $\c{d}{kjm}$ are
\begin{eqnarray}
\a{d}{kjm} &=& (-1)^k \sqrt{\tfrac{4\pi k!}{(k+j+1)!!(k-j)!!}}\bc{d-2}{k}
\big(\Yrjmconj{k}{jm}\big)^{a_1\ldots a_k}
a_w{}_\eff^{(d)a_1\ldots a_k0\ldots0} \ ,
\notag\\
\c{d}{kjm} &=& (-1)^k \sqrt{\tfrac{4\pi k!}{(k+j+1)!!(k-j)!!}}\bc{d-2}{k}
\big(\Yrjmconj{k}{jm}\big)^{a_1\ldots a_k}
c_w{}_\eff^{(d)a_1\ldots a_k0\ldots0} \ ,
\label{ac_rel}
\end{eqnarray}
where $\bc{m}{n}$ are binomial coefficients,
and $\Yrjm{k}{jm}$ are the orthonormal
spherical-harmonic tensors recently derived in Ref.\ \cite{Yrjm}.
We use Latin indices $a,b,\ldots$
to indicate the restriction
to spatial dimensions.
In many cases,
a nonrelativistic approximation is warranted,
leading to a third version,
\begin{equation}
\de h_w = \sum_{kjm} |\pvec_w|^k\,
\syjm{}{jm}(\phat_w)\, \big(\anr{kjm} - \cnr{kjm}\big) \ ,
\label{hnr}
\end{equation}
where the nonrelativistic spherical coefficients
for Lorentz violation are
\begin{eqnarray}
\anr{kjm} &=& \sum_{dl} \bc{(d-3-k+2l)/2}{l}
M_w^{d-3-k} \a{d}{(k-2l)jm}\ ,
\notag \\
\cnr{kjm} &=& \sum_{dl} \bc{(d-3-k+2l)/2}{l}
M_w^{d-3-k} \c{d}{(k-2l)jm}\ , 
\end{eqnarray}
with $k-j = \text{even}\geq 0$.
All spherical coefficients,
including the composite coefficients derived below,
obey the complex conjugation relation
$\K_{jm}^* = (-1)^m\K_{j(-m)}$.

Next, we envision a small but macroscopic volume of
matter of mass $M$ containing
a large number of electrons, protons, and neutrons.
We write the Lorentz-violating
change in the hamiltonian for the volume as
$\de H = \de H_e + \de H_p + \de H_n$,
where $\de H_e$, $\de H_p$, and $\de H_n$
represent the total free-particle hamiltonians for
electrons, protons, and neutrons, respectively.
Each can be written as
\begin{equation}
\de H_w = N_w \sum_{kjm}
\bigvev{|\pvec_w|^k\, \syjm{}{jm}(\phat_w)}_w\,
\big(\anr{kjm} - \cnr{kjm}\big) \ ,
\label{dHw}
\end{equation}
where $N_w$ is the number of particles of species $w$
in the volume,
and brackets $\vev{}_w$ indicate
the average over all the particles of that species.
We then split the bulk motion from the internal motion of the particles.
Let $\pvec$ be the total momentum,
which is conjugate to the center-of-mass position $\xvec$.
Then $\pvec_w' = \pvec_w - \frac{M_w}{M}\pvec$
is the conventional momentum of a particle
in the center of mass frame.
Normally $\pvec/M$ is
the velocity of the center of mass
and $\vvec'_w = \pvec'_w/M_w$
is the velocity of a particle relative
to the center of mass,
but this may no longer be true in
the Lorentz-violating case.
However, in leading-order calculations,
we can assume the usual relations
in the Lorentz-violating contributions
to the hamiltonian since corrections to the velocity
would produce higher-order effects.

The average in \rf{dHw} can be written as
\begin{eqnarray}
\bigvev{|\pvec_w|^k\, \syjm{}{jm}(\phat_w)}_w
&=&
\sqrt{\tfrac{(k+j+1)!!(k-j)!!}{4\pi k!}}\Yrjm{k}{jm}\cdot
\bigvev{\pvec_w^{\odot k}}_w
\notag\\
&=&
\sqrt{\tfrac{(k+j+1)!!(k-j)!!}{4\pi k!}}
\Yrjm{k}{jm}\cdot
\sum_q\bc{k}{q}
M_w^{k-q} M^{q-k}
\bigvev{\pvec_w'^{\odot q}}_w \odot \pvec^{\odot(k-q)}\ ,
\end{eqnarray}
where $\odot$ represents the symmetric tensor product,
and $\Yrjm{k}{jm}$ are the rank-$k$ spherical-harmonic tensors \cite{Yrjm}.
The product $\pvec_w'^{\odot q}$
can be expanded in spherical-harmonic tensors, giving
\begin{equation}
\pvec_w'^{\odot q} 
=\sum_{j'm'} \sqrt{\tfrac{4\pi q!}{(q+j'+1)!!(q-j')!!}}
|\pvec_w'|^q \yjm{j'm'}(\phat_w') \Yrjmconj{q}{j'm'}\ .
\end{equation}
We then make the simplifying assumption that
the internal-momentum distribution is approximately isotropic.
This implies the $j'=m'=0$ term in the sum dominates
when averaged over the particles,
yielding the approximation
$\bigvev{\pvec_w'^{\odot q}}_w
\approx \bigvev{|\pvec_w'|^q}_w \Yrjmconj{q}{00}/\sqrt{q+1}$,
which vanishes for odd values of $q$.
Replacing $q$ with $2q$,
calculation then gives
\begin{equation}
\bigvev{|\pvec_w|^k\, \syjm{}{jm}(\phat_w)}_w
\approx
\sum_q
\tfrac{(k+j+1)!!(k-j)!!}{(2q+1)!(k-2q+j+1)!!(k-2q-j)!!}
M_w^{k-2q}
M^{2q-k}
\bigvev{|\pvec_w'|^{2q}}_w 
|\pvec|^{(k-2q)}\yjm{jm}(\phat) \ ,
\end{equation}
where the sum is limited to integers $q$
for which the arguments of all factorials
are nonnegative.
The Lorentz-violating hamiltonian for
the small macroscopic volume
of matter then reads
\begin{equation}
\de H \approx
-\sum_{kjm} M^{1-k} |\pvec|^k
\syjm{}{jm}(\phat)\, \cT{kjm} \ ,
\label{dH}
\end{equation}
where we define composite coefficients for Lorentz violation
\begin{equation}
\cT{kjm} = \sum_{wq}
\tfrac{(k+2q+j+1)!!(k+2q-j)!!}{(2q+1)!(k+j+1)!!(k-j)!!}
\frac{\rh_w}{\rh}
M_w^{k-1}
\bigvev{|\pvec_w'|^{2q}}_w 
\big(\cnr{(k+2q)jm} - \anr{(k+2q)jm}\big) \ .
\label{cT1}
\end{equation}
The $q$ index sums over all nonnegative integers,
the index $k$ is restricted to nonnegative values,
and $j=k, k-2,k-4,\ldots \geq 0$.

The macroscopic coefficients for Lorentz violation in \rf{cT1}
are the dimensionless combinations of the particle coefficients
that affect macroscopic matter at leading order.
They are defined so that they depend
on the material's particle content
and not the size or shape of the object.
The macroscopic coefficients are controlled by
the mass fractions $\rh_w/\rh$,
where $\rh_w$ is the mass density for each species
and $\rh=\sum_w\rh_w$ is the total density.
Materials with different particle content
have different $\cT{kjm}$ coefficients,
so constraints on Lorentz violation in
multiple different forms of matter
could be combined to separately
constrain violations in
electrons, protons, and neutrons.

Writing the $\cT{kjm}$ in terms of
the relativistic coefficients,
\begin{align}
\cT{kjm}
&= \sum_{wdql}
\bc{(d-3-k-2q+2l)/2}{l}
\tfrac{(k+2q+j+1)!!(k+2q-j)!!}{(2q+1)!(k+j+1)!!(k-j)!!}
\frac{\rh_w}{\rh}
M_w^{d-4-2q}
\notag\\
&\qquad\quad
\times
\bigvev{|\pvec_w'|^{2q}}_w \,
\big(\c{d}{(k+2q-2l)jm}-\a{d}{(k+2q-2l)jm}\big) \ ,
\label{cT2}
\end{align}
we note that both $\a{d}{kjm}$ and $\c{d}{kjm}$ coefficients
for all dimensions $d$ contribute at each $k$,
so the effects of CPT-even and CPT-odd
Lorentz violation cannot be disentangled using only normal matter.
We therefore define a single set
of effective $\cT{kjm}$ coefficients for matter.
These will, however, differ from the macroscopic coefficients
for antimatter due to CPT violation.
Also note that for fixed $d$,
the $\cT{kjm}$ combinations
depend on the internal velocity $\vvec_w' = \pvec_w'/M_w$
through $\bigvev{|\vvec_w'|^{2q}}_w$.
Under normal circumstances,
the electron internal energy is less than a keV.
The internal energies of protons and neutrons
is typically on the order of 10 MeV.
So while the nonrelativistic internal motions
contribute to macroscopic hamiltonian,
their effects for fixed $d$ are highly suppressed relative
the $\pvec_w'$-independent terms with $q=0$.
The suppressed terms could, however,
be used to access different combinations of coefficients.
Ignoring the contributions for the internal velocities
gives the simplifying approximation
\begin{equation}
\cT{kjm}
\approx
\sum_{wdl}
\bc{(d-3-k+2l)/2}{l}
\frac{\rh_w}{\rh} M_w^{d-4}
\big(\c{d}{(k-2l)jm} - \a{d}{(k-2l)jm}\big) \ .
\label{cT3}
\end{equation}
Combined with \rf{dH},
this connects the underlying coefficients
for Lorentz violation to
the dynamics of macroscopic matter.
In ordinary neutral matter made
of atoms with atomic number $Z$
and atomic mass $M_a$,
The composite coefficients simplify to
\begin{align}
\cT{kjm}
&\approx
\sum_{dl}
\bc{(d-3-k+2l)/2}{l}\bigg[
\frac{Z}{M_a} M_n^{d-3}
\big(\ct{d}{(k-2l)jm} - \at{d}{(k-2l)jm}\big)
\notag\\
&\qquad\qquad\qquad\qquad\qquad\qquad
+ M_n^{d-4}
\big(\cw{n}{d}{(k-2l)jm} - \aw{n}{d}{(k-2l)jm}\big)
\bigg] \ ,
\label{cT4}
\end{align}
where we define coefficient combinations
\begin{equation}
\ct{d}{kjm} =
- \frac{M_p+M_e}{M_n}\cw{n}{d}{kjm}
+\bigg(\frac{M_p}{M_n}\bigg)^{d-3} \cw{p}{d}{kjm}
+ \bigg(\frac{M_e}{M_n}\bigg)^{d-3} \cw{e}{d}{kjm}
\label{ct}
\end{equation}
for even $d$,
with a similar expression for odd-$d$ $\at{d}{kjm}$ coefficients.
These coefficient combinations lead to different effects
in different types of matter and can be tested
in experiments comparing test bodies made of different elements.
These include the equivalence-principle experiments
discussed in Section \ref{WEP}.
In contrast, the remaining parts of $\cT{kjm}$
involving the neutron coefficients for Lorentz violation
produce identical affects in all matter
and are not testable through matter-comparison experiments.
Note that the above combinations mirror ones arising
in studies of matter-gravity coupling in the SME \cite{kt11}.
A partial match to this work is given in the next section.

Lorentz violation introduces signatures
other than composition-dependent dynamics.
Searches for these signatures in ordinary matter
are less dependent on precise makeup of the mass.
In this case, it may suffice to assume roughly equal numbers
of neutrons, protons, and electron,
which leads to the approximation
\begin{equation}
\cT{kjm}
\approx
\sum_{dl}
\bc{(d-3-k+2l)/2}{l}\half
M_n^{d-4} \big(\ctt{d}{(k-2l)jm} - \att{d}{(k-2l)jm}\big)\ ,
\label{cT5}
\end{equation}
where
\begin{equation}
\ctt{d}{kjm} =
\cw{n}{d}{kjm} + \cw{p}{d}{kjm}
+ \bigg(\frac{M_e}{M_n}\bigg)^{d-3} \cw{e}{d}{kjm}\ ,
\label{ctt}
\end{equation}
with a similar expression for $\at{d}{kjm}$.
Note that this approximation also
neglects the difference in the proton and neutron masses.

A cartesian version of the macroscopic coefficients
for Lorentz violation may be convenient in some applications.
Defining
\begin{equation}
\big(\cT{k}\big)^{a_1\ldots a_k}
= \sum_{jm}\sqrt{\tfrac{(k+j+1)!!(k-j)!!}{4\pi k!}}
\cT{kjm} \big(\Yrjm{k}{jm}\big)^{a_1\ldots a_k} \ ,
\end{equation}
the Lorentz-violating hamiltonian becomes
\begin{equation}
\de H = -\sum_k M^{1-k} \big(\cT{k}\big)^{a_1\ldots a_k} p^{a_1}\ldots p^{a_k}\ .
\end{equation}
Inverting the relation gives
\begin{equation}
\cT{kjm} =
\sqrt{\tfrac{4\pi k!}{(k+j+1)!!(k-j)!!}}
\big(\Yrjmconj{k}{jm}\big)^{a_1\ldots a_k}
\big(\cT{k}\big)^{a_1\ldots a_k}\ .
\end{equation}
The
$\big(\cT{k}\big)^{a_1\ldots a_k}$
tensors are real and totally symmetric.
The $k$ index on composite coefficients
is restricted to nonnegative integers,
and the angular-momentum indices obey
$j=k, k-2,k-4,\ldots \geq 0$ and $|m|\leq j$.

\subsection{Equations of motion}
\label{EOM}

The Lorentz-violating hamiltonian $\de H$
depends on the center-of-mass momentum $\pvec$,
but is independent of the center-of-mass position $\xvec$.
This implies that the net force,
defined as the rate of change in the canonical momentum,
is unchanged by Lorentz violation:
$\Fvec = \prt_t \pvec = -\del_{\!x} \de H$.
The Lorentz violation enters through
the altered relationship between the momentum $\pvec$
and the center-of-mass velocity:
$\dot{\xvec} = \del_{\!p} H
= \pvec/M + \del_{\!p}\, \de H$.
Combining the two Hamilton's equations,
we arrive at a modified Newton's second law,
\begin{equation}
M\ddot{\xvec} = (1 - C) \cdot \Fvec \ ,
\label{N2}
\end{equation}
where Lorentz violation is governed by
the symmetric dimensionless tensor
\begin{eqnarray}
C^{ab} &=& - M \frac{\prt^2\de H}{\prt p^{a}\prt p^{b}}
\notag\\
&=& 
\sum_k  k(k-1)\big(\cT{k}\big)^{abc_1\ldots c_{k-2}}
v^{c_1}\ldots v^{c_{k-2}} \ .
\label{C}
\end{eqnarray}
We write this in terms of the conventional
velocity $\vvec = \pvec/M$ for convenience.
Note that $\vvec$ can be taken as $\dot{\xvec}$
in leading-order calculations.
We then find a Lorentz-violating force
$\de \Fvec \approx -C \cdot\Fvec$
that depends on the velocity $\vvec$
and the conventional force $\Fvec$.
Alternatively,
we can write the equations of motion as
$M(1+C)\cdot \ddot{\xvec} = \Fvec$,
where the effects of Lorentz violation
can be viewed as a velocity-dependent
anisotropic mass matrix $M(1+C)$.
This generalizes the modifications
from $d=4$ violations found in 
Refs.\ \cite{bertschinger}
and \cite{clyburn}.
  
The velocity-dependent $C$ tensor can be
expanded in spin-weighted spherical harmonics.
This is done by expanding the tensor
in the helicity basis \cite{km09}:
\begin{eqnarray}
\Er &=& \ER = \vhat
= \sin\th\cos\ph\,\Ex
+ \sin\th\sin\ph\,\Ey
+ \cos\th\,\Ez ,
\notag \\
\Epm &=& \EMP
= \tfrac{1}{\sqrt{2}}(\Eth \pm i\Eph) \ .
\label{evecs}
\end{eqnarray}
The velocity direction $\vhat = \vvec/|\vvec|$
defines the ``radial'' direction,
and $\Eth$ and $\Eph$
are the usual unit vectors associated with
spherical-coordinate angles $\th$ and $\ph$.
The components $C_{ab} = \ehat_a\cdot C \cdot\ehat_b$
in the helicity basis are spin-weighted functions
and can be expanded in spin-weighted spherical harmonics $\syjm{s}{jm}$.
The result is
\begin{eqnarray}
C_{rr} &=& \sum_{kjm} k(k-1)
|\vvec|^{k-2}\, \syjm{0}{jm}(\vhat)\, \cT{kjm}\ ,
\notag \\
C_{+-} &=& \sum_{kjm}\big(k-\half j(j+1)\big)
|\vvec|^{k-2} \syjm{0}{jm}(\vhat)\, \cT{kjm}\ ,
\notag \\
C_{r\pm} &=& \sum_{kjm}
(\mp) (k-1) \sqrt{\tfrac{j(j+1)}{2}}
|\vvec|^{k-2} \syjm{\pm1}{jm}(\vhat)\, \cT{kjm}\ ,
\notag \\
C_{\pm\pm} &=& \sum_{kjm}
\half\sqrt{(j-1)j(j+1)(j+2)}
|\vvec|^{k-2} \syjm{\pm2}{jm}(\vhat)\, \cT{kjm}\ .
\end{eqnarray}
All of the objects appearing
in these expressions are dimensionless.

We note that only composite coefficients
with $k\geq 2$ affect the macroscopic dynamics
at leading-order.
For these coefficients,
the effects are proportional to $|\vvec|^{k-2}$.
Since $\vvec$ is the velocity relative to the speed of light,
the Lorentz violation from $k>2$
will be highly suppressed in most applications.
Consequently,
the dominant effects are likely those
from the $k=2$ macroscopic coefficients $\cT{2}$.
In the $k=2$ restriction,
the tensor
$C^{ab} = 2 \big(\cT{2}\big)^{ab}$
is constant,
and its cartesian components
are linear combinations
of the spherical coefficients $\cT{2jm}$:
\begin{eqnarray}
C^{xx} &=&
\sqrt{\tfrac{1}{\pi}} \cT{200}
-\sqrt{\tfrac{5}{4\pi}} \cT{220}
+\sqrt{\tfrac{15}{2\pi}} \re\cT{222}
\ , \notag \\
C^{yy} &=&
\sqrt{\tfrac{1}{\pi}} \cT{200}
-\sqrt{\tfrac{5}{4\pi}} \cT{220}
-\sqrt{\tfrac{15}{2\pi}} \re\cT{222}
\ , \notag \\
C^{zz} &=&
\sqrt{\tfrac{1}{\pi}} \cT{200}
+\sqrt{\tfrac{5}{\pi}} \cT{220}
\ , \notag \\
C^{xy} &=&
-\sqrt{\tfrac{15}{2\pi}} \im\cT{222}
\ , \notag \\
C^{xz} &=&
-\sqrt{\tfrac{15}{2\pi}} \re\cT{221}
\ , \notag \\
C^{yz} &=&
\sqrt{\tfrac{15}{2\pi}} \im\cT{221}
\ .
\end{eqnarray}
The resulting velocity-independent
effects are then limited
to  $j=0$ isotropic violations
and $j=2$ quadrupole anisotropies.

The $k=2$ case can be partially mapped
onto previous analyses of composite matter
in the SME.
In particular,
Ref.\ \cite{kt11} derives the effects
of $d=3$ and $d=4$ violations
in matter and gravity,
including the matter-gravity coupling.
Dropping the violations involving gravity
and those that are cubic in velocity,
the coefficients in that work correspond to
$C^{ab} = -2\bar c^{Tab} - \bar c^{Ttt}\de^{ab}$.
This provides a map for $d=4$ coefficients in the two approaches,
which can be extended to higher-$d$ violations
through their contributions to $\big(\cT{2}\big)^{ab}$ coefficients.
This connection could, in principle,
be used to convert bounds on $\bar c^{T}$
to bounds on $d>4$ violations.
Note, however,
that this may be problematic in analyses
involving boosts of the apparatus
since the contributions from different $d$
transform differently.
Rotations and boosts of the coefficients
are described in the next section.

\subsection{Lorentz Transformations}
\label{LT}

Many tests of Lorentz invariance
search for changes in a signal
with changes in the orientation or
velocity of the apparatus.
The changes in orientation are typically due to
the daily rotation of the Earth,
but can be achieved through the use of turntables.
The changes in velocity
are usually those resulting from
the orbital motion of the Earth.
Assuming constant coefficients for Lorentz violation
in inertial frames,
the above motions produce variations
in the coefficients in noninertial
apparatus-fixed frames.
These variations lead to variations
in experimental observables,
producing potential signals of Lorentz violation.
In this section,
we review rotations of spherical coefficients
and discuss common frames used in tests
of Lorentz invariance.
We also derive the boosts of
spherical coefficients for Lorentz violation
to first order in boost velocity.

Rotations of spherical-harmonic expansion coefficients
are given by Wigner matrices.
Consider the expansion of a spin-weighted function
$f(\nhat) = \sum_{jm} f_{jm}\, \syjm{s}{jm}(\nhat)$,
where $\nhat$ is a direction unit vector
We then consider two frames whose coordinates
$\{x,y,z\}$ and $\{x',y',z'\}$ are related through
\begin{equation}
\begin{pmatrix}
  dx'\\ dy'\\ dz'
\end{pmatrix}
=
\begin{pmatrix}
  \cos\al&-\sin\al&0\\
  \sin\al&\cos\al&0\\
  0&0&1
\end{pmatrix}
\begin{pmatrix}
  \cos\be& 0&\sin\be\\
  0&1&0\\
  -\sin\be& 0&\cos\be
\end{pmatrix}
\begin{pmatrix}
  \cos\ga&-\sin\ga&0\\
  \sin\ga&\cos\ga&0\\
  0&0&1
\end{pmatrix}
\begin{pmatrix}
  dx\\ dy\\ dz
\end{pmatrix} \ .
\end{equation}
The connection between spherical-harmonic components of $f$
in these two frames is given
\begin{eqnarray}
f'_{jm} &=& \sum_{m'} D^{(j)}_{mm'}(\al,\be,\ga) f_{jm'}
\notag \\
&=& \sum_{m'} e^{-im\al} e^{-im'\ga}\, d^{(j)}_{mm'}(\be) f_{jm'}\ ,
\end{eqnarray}
where $D^{(j)}_{mm'}(\al,\be,\ga)$ 
and $d^{(j)}_{mm'}(\be) = D^{(j)}_{mm'}(0,\be,0)$
are Wigner matrices.

By convention,
tests involving the SME report results
in a Sun-centered celestial equatorial
inertial reference frame with spacetime coordinates $\{T,X,Y,Z\}$.
The $Z$ axis points along the Earth's rotation axis,
$X$ points towards the vernal equinox,
and $Y$ completes the system.
The standard time $T$ is define so that $T=0$
at vernal equinox in the year 2000.
A standard noninertial laboratory frame $\{t,x,y,z\}$ is defined
with $x$ and $y$ horizontal and $z$ pointing vertically up.
The $x$ axis points at an angle $\vp$ east of south.
In order to incorporate boosts,
we define an intermediate Earth-centered frame
with coordinates $\{T',X',Y',Z'\}$
that is boosted but not rotated
relative to the Sun frame.
The rotation relating the lab frame and the Earth frame is
\begin{equation}
\begin{pmatrix}
  dX'\\ dY'\\ dZ'
\end{pmatrix}
=
\begin{pmatrix}
  \cos\al&-\sin\al&0\\
  \sin\al&\cos\al&0\\
  0&0&1
\end{pmatrix}
\begin{pmatrix}
  \cos\ch& 0&\sin\ch\\
  0&1&0\\
  -\sin\ch& 0&\cos\ch
\end{pmatrix}
\begin{pmatrix}
  \cos\vp&-\sin\vp&0\\
  \sin\vp&\cos\vp&0\\
  0&0&1
\end{pmatrix}
\begin{pmatrix}
  dx\\ dy\\ dz
\end{pmatrix} \ ,
\end{equation}
where $\ch$ is the colatitude of the laboratory,
and $\al$ is the right ascension of the laboratory zenith.
The transformation of spherical coefficients leads to
\begin{equation}
f^\text{lab}_{jm}
= \sum_{m'} D^{(j)}_{mm'}(-\vp,-\ch,-\al) f^\text{Earth}_{jm'} \ .
\label{rot}
\end{equation}
The right ascension increases at Earth's sidereal rate
$\dot\al \approx \om_\oplus = 2\pi/\text{23 hr 56 min}$,
producing a sidereal variation in laboratory-frame coefficients.
Horizontal turntables would give variations in $\vp$ as well.

Ignoring boosts,
the Earth-frame and Sun-frame coefficients
are the same up to a time-dependent translation.
We next determine the relationship
between these frames to first order in
the boost velocity $\bevec$.
The transformation depends on the tensor structure of
the underlying coefficients for Lorentz violation.
Here, we focus on the relativistic spherical
$\a{d}{kjm}$ and $\c{d}{kjm}$
coefficients in \rf{ac_rel}.
The derivation is the same for both of these sets of coefficients,
so we show the calculation for $\a{d}{kjm}$ only.

Inverting relationship \rf{ac_rel}
between the spherical and cartesian representations
of the coefficients for Lorentz violation,
we can write
\begin{equation}
a_w{}_\eff^{(d)a_1\ldots a_k0\ldots0}
=
(-1)^k \tfrac{(d-2-k)!}{(d-2)!}
\sum_{jm}
\sqrt{\tfrac{k!(k+j+1)!!(k-j)!!}{4\pi}}
\a{d}{kjm}
\big(\Yrjm{k}{jm}\big)^{a_1\ldots a_k}\ .
\label{aYexp}
\end{equation}
Consider a ``primed'' frame that is 
moving with small boost velocity $\bevec$ relative
to a second ``unprimed'' frame.
The difference between 
the cartesian coefficients
$\de a_w{}_\eff^{(d)}
=a'_w{}_\eff^{(d)}-
a_w{}_\eff^{(d)}$
in the two frames is then given by
\begin{equation}
\de a_w{}_\eff^{(d)a_1\ldots a_k0\ldots0} =
-\tfrac{1}{(k-1)!} a_w{}_\eff^{(d)0\ldots0(a_1\ldots a_{k-1}}\be^{a_k)} 
-(d-2-k) a_w{}_\eff^{(d)0\ldots0a_1\ldots a_kb}\be^{b} \ .
\label{da_cart}
\end{equation}
Using \rf{aYexp},
we can expand the right-hand side
of the above expression in terms
of spherical coefficients.
Expanding the boost velocity in spherical-harmonic tensors,
$\be^a = \sqrt{\frac{4\pi}{3}}\be \sum_m \yjm{1m}(\behat) (\Yjm{1m}^*)^a$,
and using the product identity derived in the Appendix,
we can expand \rf{da_cart} in spherical-harmonic tensors.
Combining the result with \rf{ac_rel}
gives the change in the spherical coefficients
for Lorentz violation.
The result is the first-order boost
of the spherical coefficients
\begin{equation}
\de\a{d}{kjm}
=
\sum_{k'j'm'm''}
\G{d}{kjm}{k'j'm'}{m''}\,
\sqrt{\tfrac{4\pi}{3}}\,
\be
\yjm{1m''}(\behat)\, 
\a{d}{k'j'm'}\ ,
\end{equation}
where $\be = |\bevec|$, $\behat=\bevec/\be$, 
and
\begin{equation}
\G{d}{kjm}{k'j'm'}{m''} =
\sqrt{\tfrac{(k'-j')!!(k'+j'+1)!!}{(k-j)!!(k+j+1)!!}}\
\begin{cases}
(-1)^{m''}
(d-1-k)
\sqrt{k}\
\A{1k'}{1(-m'')j'm'jm}\ ,
\quad& k'=k-1\ , \\
\sqrt{k'}\
\A{1k}{1m''jmj'm'}\ ,
\quad& k'=k+1\ , \\
0\ , \quad& \text{otherwise,}
\end{cases}
\label{Gamma}
\end{equation}
in terms of the
$\A{\vr_1\vr_2}{j_1m_1j_2m_2JM}$
constants in \rf{Acoeffs}.

The velocity of the Earth in the Sun frame
is approximately given by
$(\be_X,\be_Y,\be_Z)
\approx \be(
\sin\Om_\oplus T,
-\cos\et \cos\Om_\oplus T,
-\sin\et \cos\Om_\oplus T)$,
where the orbital speed is $\be\approx 9.9\times10^{-5}$,
$\et\approx 23.4^\circ$ is the inclination of the orbit,
and $\Om_\oplus=2\pi/\text{year}$
is the orbital frequency.
With this, we can write
$\sqrt{\tfrac{4\pi}{3}}\, \be \yjm{1m}(\behat)
= \sum_{m'} B_{mm'} e^{im'\Om_\oplus T}$,
where the nonzero $B_{mm'}$ constants are
\begin{equation}
B_{(\pm1)(\pm1)} = \frac{i\be}{\sqrt2}\cos^2\frac\et2\ ,\quad
B_{(\pm1)(\mp1)} = -\frac{i\be}{\sqrt2}\sin^2\frac\et2\ ,\quad
B_{0(\pm1)} = -\frac{\be}{2}\sin\et\ .
\label{Bs}
\end{equation}
Combining the boost between the Sun and Earth frame
with the rotation between the Earth and laboratory frame,
we find the coefficients transform according to
\begin{eqnarray}
\alab{d}{kj\mt} &=&
\sum_{\ms} e^{i\mt\vp + i\ms\al} d^{(j)}_{\mt\ms}(-\ch) \asun{d}{kj\ms}
\notag \\
&&+
\sum_{\ms\ma} e^{i\mt\vp+i\ms\al+i\ma\Om_\oplus T}
d^{(j)}_{\mt\ms}(-\ch)
\sum_{k'j'm'm''}
\G{d}{kj\ms}{k'j'm'}{m''}\,
B_{m''\ma}
\asun{d}{k'j'm'}\ .
\label{lt}
\end{eqnarray}
The indices $\mt$, $\ms$, and $\ma$
are the harmonic numbers for variations
at the turntable rotation rate $\dot\vp$,
the sidereal rate $\dot\al$,
and the annual rate $\Om_\oplus$,
respectively.
The $\c{d}{kjm}$ coefficients in
the laboratory and Sun frames
obey the same relationship.

\section{Applications}
\label{APPLICATIONS}

\subsection{Tests of the Equivalence Principle}
\label{WEP}

The modified equations of motion \rf{N2}
imply that the acceleration of a mass under
the influence of gravity depends on its particle content,
giving an apparent violation of the weak equivalence principle (WEP).
Consequently, tests of the WEP may be adapted
for searches for nonminimal Lorentz violations in particle sectors
of the SME.
Most WEP tests involving macroscopic masses fall into two main categories,
free-fall experiments and torsion-pendulum experiments.
We consider each of these in turn.

In first class of experiment,
the vertical free-fall acceleration of two test masses
made of different material are compared.
In the experiments performed to date,
each mass is released from rest in a vacuum chamber.
Accelerations are measured using lasers
and corner-cube reflectors
attached to the masses \cite{ff86,ff87,ff89,ff90,ff92,ff96}.

We work in the standard laboratory frame
described in Section \ref{LT}
and assume the gravitational field
$\gvec = - g\Ez$
is approximately uniform.
The leading-order change in the acceleration is given by
$\de\ddot\xvec = g\, C\cdot\Ez$,
where the $C$ tensor is a function 
of the unperturbed velocity
$\vvec = -g t\Ez$.
For fixed $k\geq 2$,
we can integrate twice to get
the change in position after time $t$,
$\de x^l = g^{-1}(-gt)^k  \big(\cTlab{k}\big)^{lz\ldots z}$,
in terms of the laboratory-frame cartesian
composite coefficients for Lorentz violation.
In principle, free-fall experiments
could search for $\sim t^k$ displacements from
the velocity-dependent $k>2$ violations.
These effects, however, are highly suppressed by
the small velocities.
For the velocity-independent $k=2$ cases,
the vertical acceleration is constant
and given by
\begin{equation}
\de\ddot z/g = 2(\cTlab{2})^{zz} =
\sqrt{\tfrac{1}{\pi}} \cTlab{200} +\sqrt{\tfrac{5}{\pi}} \cTlab{220}\ .
\end{equation}
This includes the isotropic $\cTlab{200}$
and the quadrupole $\cTlab{220}$.
The quadrupole violations lead to sidereal variations,
which could be sought in future analyses.
However,
in order to understand the reach of these experiments,
we will ignore boosts and focus on isotropic violations
$\cTlab{200}\approx \cT{200}$.
In this particular limit,
the acceleration is proportional to the gravitational field,
$\ddot\xvec = (1-\sqrt{\tfrac{1}{\pi}} \cT{200}) \gvec$.
So isotropic Lorentz violation mimics a difference in
the inertial mass $M_I$ and gravitational mass $M_g$.
This type of behavior is traditionally characterized using
the E\"otv\"os parameter,
defined as the difference in free-fall acceleration $\De a$
divided by the average  $\bar a$ for the two test masses.
In terms of the masses,
this gives $\et = \De a/\bar a = \De r/\bar r$,
where $r=M_g/M_I$.
In the present context,
we find get an effective Lorentz-violating E\"otv\"os parameter
\begin{equation}
\et_\text{LV}
= -\sqrt{\tfrac{1}{\pi}} \De\cT{200}
=\De \bigg(\frac{Z}{M_a}\bigg) M_n^{d-3}
\begin{cases}
  -\ct{d}{200} - \tfrac{d-3}{2}\ct{d}{000} \ , & d=\text{even},\\
  \at{d}{200} + \tfrac{d-3}{2}\at{d}{000} \ , & d=\text{odd},\\
\end{cases}
\label{eotvos}
\end{equation}
for fixed dimension $d$.
The differences in the test bodies enter
through the difference $\De(Z/M_a)$
in the ratio of the atomic number and
atom mass.
Note that the simple correspondence \rf{eotvos}
breaks down in more general cases
where there are accelerations perpendicular to $\gvec$
and velocity-dependent accelerations.

\begin{table}
\begin{tabular}{c|c|c|c}
  coefficients
  & free fall 
  & MICROSCOPE
  & torsion pendulum\\
  \hline
  $\ct{4}{200}+\tfrac{1}{2} \ct{4}{000}$ &
  $(-5\pm 12)\times 10^{-9}$ &
  $(3\pm 27\pm 27)\times 10^{-14}$ &
  $(3\pm 7)\times 10^{-12}\text{ GeV}$
  \\
  $\at{5}{200}+ \at{5}{000}$ &
  $(5\pm 13)\times 10^{-9}\text{ GeV}^{-1}$ &
  $(-3\pm 29\pm 29)\times 10^{-14}\text{ GeV}^{-1}$ &
  $(3\pm 8)\times 10^{-12}\text{ GeV}^{-1}$
  \\
  $\ct{6}{200}+\tfrac{3}{2} \ct{6}{000}$ &
  $(-6\pm 13)\times 10^{-9}\text{ GeV}^{-2}$ &
  $(3\pm 30\pm 30)\times 10^{-14}\text{ GeV}^{-2}$ &
  $(3\pm 8)\times 10^{-12}\text{ GeV}^{-2}$
  \\
  $\at{7}{200}+ 2\at{7}{000}$ &
  $(6\pm 14)\times 10^{-9}\text{ GeV}^{-3}$ &
  $(-4\pm 32\pm 32)\times 10^{-14}\text{ GeV}^{-3}$ &
  $(4\pm 9)\times 10^{-12}\text{ GeV}^{-3}$
  \\
  $\ct{8}{200}+\tfrac{5}{2} \ct{8}{000}$ &
  $(-6\pm 15)\times 10^{-9}\text{ GeV}^{-4}$ &
  $(4\pm 34\pm 34)\times 10^{-14}\text{ GeV}^{-4}$ &
  $(4\pm 10)\times 10^{-12}\text{ GeV}^{-4}$
\end{tabular}
\caption{
  \label{eptable}
  Limits on isotropic SME coefficients from tests of
  the equivalence principle.
  The first column gives the coefficient combinations.
  The second column contains the combined constraints
  from ground-based free-fall experiments.
  The third column lists constraints from the space-based
  MICROSCOPE experiment.
  The last column gives the combined constraints
  from torsion-pendulum experiments.
}
\end{table}

Using the above,
we can translate published constraints
on $\et$ to measurements
on the dimensionless isotropic-coefficient combinations
$-M_n^{d-4}\big( \ct{d}{200}+\tfrac{d-3}{2} \ct{d}{000}\big)$
for even $d$ and
$M_n^{d-4}\big( \at{d}{200}+\tfrac{d-3}{2} \at{d}{000}\big)$
for odd $d$.
A number of different ground-based experiments
have compared the free-fall of different materials
in the Earth's gravitational field at the level of $\et \sim 10^{-9}$.
Translating these to constraints on
the above combinations of SME coefficients, we find
$(3 \pm  13)\times 10^{-9}$ using copper and uranium \cite{ff87},
$(21 \pm 60)\times 10^{-9}$ using aluminum and beryllium \cite{ff90},
$(-10\pm 58)\times 10^{-9}$ using aluminum and copper \cite{ff90},
$(17\pm 131)\times 10^{-9}$ using aluminum and carbon \cite{ff90},
$(21\pm 53\pm 74)\times 10^{-9}$ using aluminum and copper \cite{ff96},
and
$(23\pm 30\pm 33)\times 10^{-9}$ copper and tungsten \cite{ff96}.
We combine these to produce a ``best-fit'' ground-based measurement
and then translate this to constraints on
the isotropic-coefficient combinations.
The results up to $d=8$ are given the second column Table \ref{eptable}.

Similar tests of WEP in space \cite{step,gg,microscope}
could also be used to search for Lorentz violation.
For example,
the T-SAGE instrument aboard the MICROSCOPE satellite
has placed a constraint of
$\et = (-1\pm 9\pm 9)\times 10^{-15}$
on the difference between the free-fall accelerations
of titanium and platinum \cite{microscope}.
The resulting constraint on isotropic Lorentz violations
are included in the third column of Table \ref{eptable},
demonstrating that sensitivities to Lorentz violation
on the order of $10^{-13}\text{ GeV}^{4-d}$ are possible.

The classic E\"otv\"os experiment \cite{eotvos}
and its descendants represent another
class of WEP tests based on torsion pendulums.
In the prototypical experiment,
two tests masses of different composition
are attached to the ends of a rod hanging from a horizontal fiber.
A difference in the gravitational
and inertial masses would lead to an imbalance
in the horizontal components of the gravitational
and centrifugal forces,
leading to a net torque about the fiber.
Modern versions achieve high sensitive by seeking modulated signals
due to the changing field from the Sun over the day \cite{Roll,Braginskii}
or by rotating the apparatus in the laboratory \cite{Adelberger,Su,Schlamminger}.

The modified Newton's law for a suspended test mass can be written as
$M(1+C)\cdot \ddot\xvec = M\gvec + \fvec$,
where $\fvec$ is the net constraint force.
While the general modification acts as an effective anisotropic
and velocity-dependent inertial-mass matrix $M(1+C)$,
the $k=2$ isotropic limit gives the same Lorentz-violating
E\"otv\"os parameter as above.
We again use this limit to estimate potential sensitivities
in these experiments.
As in the free-fall experiments,
we convert measurements of $\et$ to constraints on isotropic coefficient combinations,
giving
$(-16\pm 13)\times 10^{-11}$ using gold and aluminum \cite{Roll},
$(0.7\pm 1.0)\times 10^{-11}$ using platinum and aluminum \cite{Braginskii},
$(-15\pm 77)\times 10^{-11}$ using beryllium and copper \cite{Adelberger},
$(15 \pm 19)\times 10^{-11}$ using beryllium and copper \cite{Su},
$(0.5\pm 7.6)\times 10^{-11}$ using beryllium and aluminum \cite{Su},
and
$(-0.2\pm 1.2)\times 10^{-11}$ using beryllium and titanium \cite{Schlamminger}.
The resulting combined constraints on
the dimension $d$ isotropic coefficients
are given in the last column of Table \ref{eptable}.

Anisotropic violations could also be tested in these experiments.
The daily rotation of the laboratory in ground-based experiments
will lead to variations at multiples of the sidereal frequency.
Signals in space-based experiments will arise at harmonics
of the satellite rotation rate.
Boosts will introduce additional frequencies in the variations of the signal,
including the annual frequency due to the motion of the Earth around the Sun.
A search for these types of variations in MICROSCOPE data
was recently carried out \cite{microscopeLV},
where $d=3$ and $d=4$ violations in the matter-gravity
couplings of the SME are constrained down to the expected
$10^{-13}\text{ GeV}^{4-d}$ range.
A similar analysis could be used to constrain
higher-order Lorentz violation.

Future space-based tests include STEP \cite{step}
and GG \cite{gg},
which could reach sensitivities two or three orders
of magnitude better than MICROSCOPE.
Other promising opportunities for future studies
include experiments utilizing
drop towers \cite{ffdt},
balloons \cite{ffballoon},
bouncing masses \cite{ffbounce},
and
sounding rockets \cite{ffsr}.

\subsection{Orbits}
\label{ORBITS}

In this section,
we consider the effects of Lorentz violation on
a satellite in a gravitational orbit around a larger body.
For simplicity,
we'll restrict attention to approximately circular orbits
and neglect effects depending on both eccentricity
and coefficients for Lorentz violation.
We work in a fixed orbit-centered frame
with the $z$ axis along the orbital axis
and $z=0$ in the orbital plane of the Lorentz-invariant limit.
We also use an orbit time $t$
defined so that the satellite velocity
is along the $x$ direction at $t=0$.
The position of the satellite can be described
using cylindrical coordinates $\{\rh,\vp,z\}$
in the orbit frame.
We denote the corresponding basis unit vectors
as $\{\Erh,\Evp,\Ez\}$.

At leading order,
the acceleration of the satellite is given by
$\ddot\xvec = \gvec - C\cdot\gvec$,
where
$\gvec = -R^3\om^2|\xvec|^{-3} \xvec$
is the gravitational field
in terms of the usual semimajor axis $R$
and the orbit angular frequency $\om$.
This leads to the change in position $\de\xvec$
of the satellite due to Lorentz violation that satisfies
$\de\ddot\xvec = -\om^2 \de\xvec
+ 3\om^2 (\Erh\cdot\de\xvec) \Erh
+ R \om^2 C \cdot\Erh$,
neglecting terms involving the small eccentricity.
The above implies the cylindrical-basis components obey 
the coupled differential equations
\begin{eqnarray}
\de\ddot x_\rh -2\om\de\dot x_\vp -3\om^2\de x_\rh 
&=& R \om^2 C_{\rh\rh} \ ,
\notag\\
\de\ddot x_\vp +2\om\de\dot x_\rh
&=& R \om^2 C_{\vp\rh} \ ,
\notag\\
\de\ddot x_z + \om^2 \de x_z
&=& R \om^2 C_{z\rh}\ .
\label{orb1}
\end{eqnarray}
The components $C_{ab} = \hat{\mbf e}_a\cdot C\cdot\hat{\mbf e}_b$
can be taken as functions of
the conventional velocity $\vvec\approx R\om\Evp$.

Since the velocity $\vvec$ is periodic
with period $2\pi/\om$,
the $C_{ab}$ components in \rf{orb1}
drive changes in the motion at harmonics of $\om$.
We characterize this driving force using the form
$C_{ab} = \sum_m C_{ab}^{[m]}e^{im\om t}$,
where $C^{[-m]}_{ab} = C^{[m]*}_{ab}$.
The $C_{ab}^{[m]}$ driving amplitudes
arise naturally out of the spherical-harmonic
expansion of the $C_{ab}$ components.
Recall that the spherical-harmonic
expansion of $C$ uses
the helicity vectors \rf{evecs}
defined with respect to
the velocity $\vvec$.
Matching to the cylindrical basis gives
$\Er = \Evp$ and 
$\Epm = -(\Ez\pm i\Erh)/\sqrt2$,
in the orbital plane.
The velocity vector points
at polar angle $\th=\pi/2$
and azimuthal angle $\ph = \om t$.
This leads to the driving amplitudes
\begin{eqnarray}
C^{[m]}_{\rh\rh} &=& \sum_{kj}
\Big[
\big(k-\tfrac12 j(j+1)\big) \syjm{0}{jm}(\Ex)
-\tfrac14\sqrt{(j-1)j(j+1)(j+2)}
\big(1 + (-1)^{j+m}\big)\syjm{+2}{jm}(\Ex)
\Big]
\notag \\ &&\qquad
\times
(R\om)^{k-2}
\cTorb{kjm}\ ,
\notag \\[6pt]
C^{[m]}_{\vp\rh} &=&
\sum_{kj} (-\tfrac{i}{2}) (k-1) \sqrt{j(j+1)}
\big(1 + (-1)^{j+m}\big)\syjm{+1}{jm}(\Ex)
(R\om)^{k-2}
\cTorb{kjm}\ ,
\notag \\[4pt]
C^{[m]}_{z\rh} &=&
\sum_{kj} (-\tfrac{i}{4})
\sqrt{(j-1)j(j+1)(j+2)} 
\big(1 - (-1)^{j+m}\big)\syjm{+2}{jm}(\Ex)
(R\om)^{k-2}
\cTorb{kjm} \ ,
\end{eqnarray}
in terms of the orbit-frame
coefficients for Lorentz violation $\cTorb{kjm}$.
These depend on spherical harmonics
for the $\Ex$ direction,
which lies at $\th=\pi/2$ and $\ph=0$.

\begin{figure}\centering
  \includegraphics[width=0.3\textwidth]{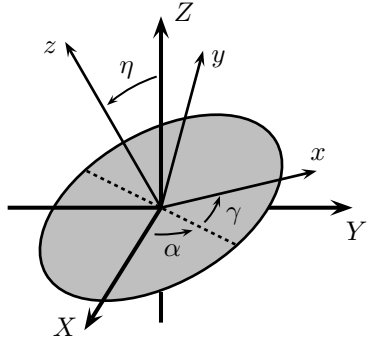}
  \caption{\label{orb_frame}
      Euler angles relating the Sun frame $\{X,Y,Z\}$
      and the orbit frame $\{x,y,z\}$.}
\end{figure}

The orbit-specific $C^{[m]}_{ab}$ coefficient combinations
determine the leading-order effects of Lorentz violation
for a particular satellite.
However, they must be connected to
the standard Sun-frame coefficients to be useful.
Ignoring boosts,
the Sun-frame coefficients $\cT{kjm}$
and the orbit-frame coefficients $\cTorb{kjm}$
are related through the rotation
\begin{eqnarray}
\cTorb{kjm} &=& \sum_{m'} D^{(j)}_{mm'}(-\ga-\tfrac\pi2,-\et,-\al+\tfrac\pi2)
\cT{kjm'}
\notag\\
&=& \sum_{m'} i^{m-m'}e^{im\ga+im'\al}d^{(j)}_{mm'}(-\et) \cT{kjm'} \ ,
\end{eqnarray}
where $\al$, $\ga$, and $\et$ are
a convenient set of Euler angles
and are illustrated in Figure \ref{orb_frame}.
The angle $\al$ is between the Sun-frame $X$ axis
and the line of nodes,
$\et$ is the inclination of the orbit relative to the $X$-$Y$ plane
and $\ga$ is the angle between the orbit-frame $x$ axis
and the line of nodes.
For our analysis,
$\ga$ is somewhat arbitrary.
It can be chosen, for example,
so that the pericenter lies on the $\Ex$ axis.
The $\al$ and $\et$ angles for the eight planets are given
in Table \ref{eulers}.

\begin{table}\centering
  \begin{tabular}{c|cccccccc}
    & Merc. & Ven. & Earth & Mars & Jup. & Sat. & Ur. & Nept. \\
    \hline
    $\al$ & $11.1^\circ$ & $8.0^\circ$ & $0^\circ$ & $3.4^\circ$ &
    $3.3^\circ$ & $6.0^\circ$ & $1.8^\circ$ & $3.5^\circ$ \\
    $\et$ & $28.5^\circ$ & $24.4^\circ$ & $23.4^\circ$ & $24.7^\circ$ &
    $23.2^\circ$ & $22.6^\circ$ & $23.7^\circ$ & $22.3^\circ$
  \end{tabular}
  \caption{\label{eulers}
    Euler angles of the planets \cite{planetangles}.}
\end{table}

To solve \rf{orb1},
we first seek force-free homogeneous solutions.
Several homogeneous solutions exist that can be
connected to conventional perturbations to circular orbits.
The general homogeneous solution is
\begin{eqnarray}
\de x_\rh^\text{hom} &=&
\de R
- R \ve \cos\om (t-t_1)
\notag \\
\de x_\vp^\text{hom} &=&
R\,\de\vp
-\tfrac32\de R\, \om t
+2R\ve\sin\om (t-t_1)\ ,
\notag \\
\de x_z^\text{hom} &=& R\et' \sin\om (t-t_2) \ .
\end{eqnarray}
The constant $\de\vp$ represents
a small translation along the orbit.
The constant $\de R$ gives a transition to
a circular orbit that is larger or smaller by $\de R$.
To see this note that while $R$ and $\om$ may change,
the combination $R^3\om^2$ is constant for Kepler orbits,
so $\de(R^3\om^2) = 3R^2\om^2\de R + 2R^3\om\de\om = 0$.
Using this,
we find that the change in position due to a change in radius is
$\de\xvec = \de R\, \Erh + R\de \om\, t\Evp
= \de R\, \Erh -\tfrac32\om t\de R\, \Evp$,
matching the above result.
This variation is only valid for sufficiently small times.
The constant $\ve$ adds small eccentricity
with a pericenter at time $t_1$.
It can be characterized using an eccentricity
vector $\vevec = \re\big(i\ve e^{-i\om t_1} (\Ex+i\Ey)\big)$,
which has magnitude $\ve$ and points to the pericenter.
The perturbations in the position can be written as
$\de x_\rh = -R \vevec\cdot\Erh,\ \de x_\vp = -2R \vevec\cdot\Evp$.
Setting $t_1$ to one quarter of the orbit period
places the pericenter on the $x$ axis.
The angle $\et'$ corresponds to a small inclination
from the orbit-frame $x$-$y$ plane
with the ascending node at time $t_2$.
Using these results,
we can distinguish between conventional perturbations of the orbit
and those caused by Lorentz violation.

For the Lorentz-violating inhomogeneous problem,
we first consider the $m=0$ and $m=1$  special cases separately
since these harmonics also appear in the homogeneous solutions.
For $m=0$, we find that $C^{[0]}_{\vp\rh} = C^{[0]}_{z\rh} = 0$.
The solution in this case is
$\de x_\rh = -\tfrac13 R C^{[0]}_{\rh\rh}$.
This gives a change in the size of the orbit
without the corresponding change in the frequency
required by Kepler's third law.
The effect mimics a small change in either
Newton's constant $G$ or the source mass.
While it may be difficult to detect,
this form of Lorentz violation could be sought,
in principle, by comparing the third-law constant $R^3\om^2$
of different satellites orbiting the same source.

For the $m=\pm1$ case, two distinct effects arise.
We first note that $C^{[1]}_{\rh\rh}=iC^{[1]}_{\vp\rh}$
and neither of $C^{[1]}_{\rh\rh}$ or $C^{[1]}_{\vp\rh}$
contain velocity-independent $k=2$ contributions.
So only $C^{[1]}_{z\rh}$ gives effects
that are unsuppressed by small orbital velocities.
Nonetheless,
the $C^{[1]}_{\rh\rh}=iC^{[1]}_{\vp\rh}$
matrix element drives a speed-dependent
change in the eccentricity.
The inhomogeneous solution can be taken as
\begin{eqnarray}
\de x_\rh &=&
-R\om t
\im\big(C^{[1]}_{\rh\rh} e^{i\om t}\big)
\notag \\
&=&
-R\om t
\big|C^{[1]}_{\rh\rh}\big|\cos\om (t-t_3)
\notag \\
\de x_\vp &=&
-2R\om t
\re\big(C^{[1]}_{\rh\rh}e^{i\om t}\big)
\notag \\
&=&
2R\om t
\big|C^{[1]}_{\rh\rh}\big|\sin\om (t-t_3)
\end{eqnarray}
where we parameterize the phase as
$C^{[1]}_{\rh\rh}
=i\big|C^{[1]}_{\rh\rh}\big|e^{-i\om t_3}$.
The above is only valid for small $t$,
but the result is an eccentricity that increases a rate of
$\dot\ve = \om\big|C^{[1]}_{\rh\rh}\big|$
with pericenter at time $t_3$.
The rate of change in the eccentricity vector is
\begin{equation}
\dot\vevec = \om\re\big(C^{[1]}_{\rh\rh}(\Ex+i\Ey)\big) \ .
\end{equation}
Note that this will add to the conventional eccentricity vector
that may point in a different direction.

The small eccentricities
of the planets lead to crude limits
on the above effect.
As an example,
consider the Earth.
The eccentricity added per orbit is $2\pi \big|C^{[1]}_{\rh\rh}\big|$,
and the Earth has made roughly
$N\approx 4.5\times 10^9$ orbits over the age of the Solar System.
Taking the Earth's eccentricity $\ve \approx 0.017$
as an upper bound on the eccentricity due to Lorentz violation,
we find the constraint
$\big|C^{[1]}_{\rh\rh}\big|
\lesssim \ve/2\pi N \simeq 6\times 10^{-13}$.
Since $k=2$ Lorentz violations do not contribute,
the dominant effects would likely be from $k=3$,
which are linear in speed.
Earth's speed is about $v\approx 10^{-4}$,
implying potential sensitivity on the order of $10^{-9}$
to $\cT{3jm}$ coefficients.
Venus, with its smaller eccentricity and shorter year,
is the only planet yielding a slightly better sensitivity.

The $C^{[1]}_{z\rh}$ coefficient combination
gives the modification
\begin{eqnarray}
\de x_z &=& -R\om t \im\big(C^{[1]}_{z\rh} e^{i\om t}\big)
\notag \\
&=& R\om t \big|C^{[1]}_{z\rh}\big|\sin\om(t-t_4) \ ,
\end{eqnarray}
where we parameterize
$C^{[1]}_{z\rh}= -\big|C^{[1]}_{z\rh}\big| e^{-i\om t_4}$.
This implies a rotation of the orbital plane
at an instantaneous rate of $\Om = \om \big|C^{[1]}_{z\rh}\big|$
about an in-plane axis pointing towards the satellite at time $t_4$.
We can account for both the rate and the direction
by defining a rotation vector
\begin{equation}
\mbf\Om = \om\im \big(C^{[1]}_{z\rh}(\Ex + i\Ey)\big)\ .
\end{equation}
The orbit axes will gradually rotate according to
$\dot{\hat{\mbf e}}_a = \mbf\Om \times \hat{\mbf e}_a$.
The resulting secular variations in the Euler angles are
\begin{eqnarray}
\dot\al &=& \om
\frac{\sin\ga\im C^{[1]}_{z\rh}+\cos\ga\re C^{[1]}_{z\rh}}{\sin\et} \ ,
\notag \\
\dot\et &=& \om \big(\cos\ga\im C^{[1]}_{z\rh}-\sin\ga\re C^{[1]}_{z\rh}\big) \ ,
\notag\\
\dot\ga &=& -\cos\et\,\dot\al \ .
\end{eqnarray}
Note that the $C^{[1]}_{z\rh}$ coefficient combination
also depends on the Euler angles.
A demonstration of the above rotation
is provided in Ref.\ \cite{clyburn},
where the effects of $d=4$ violations
on binary systems are simulated.

Assuming a small net effect,
we can approximate the total rotation of the orbital plane
after $N$ orbits as $2\pi N \big|C^{[1]}_{z\rh}\big|$.
Given that the orbital planes of the eight planets and
the Sun's equator all differ by no more than about $10^\circ$,
we take this as an upper bound on the change in inclination
for each planet over the age of the Solar System.
Mercury, with its short year, then produces the tightest constraint,
$\big|C^{[1]}_{z\rh}\big| \lesssim 1.5\times 10^{-12}$.
Unlike the change in eccentricity,
velocity-independent $k=2$ violations
contribute to the change in inclination.
So planetary orbits could constrain
$\cT{22m}$ quadrupole coefficients at the level of $10^{-12}$.
The different orientations of the different orbits could
in principle be used to access different combinations of coefficients.
However,
the small inclinations imply the effects of Lorentz violation
on all planets depend on similar combinations,
reducing the sensitivity to the additional coefficient space
accessible through a combined analysis.

Planetary ephemerides could be used to place more rigorous bounds
at levels comparable to the simple estimate given above.
For example, limits on the evolution of planetary orbits
have been shown to constrain the dimensionless
$\bar s^{\mu\nu}$ coefficients in
the pure-gravity sector of the SME down to parts in $10^{12}$
\cite{iorio,hees}.
Lunar laser ranging has also been used to test
Lorentz symmetry in gravity \cite{llr1,llr2}
and in matter-gravity couplings \cite{llr3}
down to parts in $10^{12}$.
The $\bar s^{\mu\nu}$ coefficients
produce effects similar to those found above \cite{qbk06},
and we expect similar constraints on $\cT{22m}$ coefficients.
This implies sensitivities at the level of
$\sim 10^{-12}\text{ GeV}^{4-d}$
to $\att{d}{kjm}$ and $\ctt{d}{kjm}$ SME coefficient combinations.

Binary pulsars provide another test of Lorentz invariance
in orbital dynamics \cite{qbk06,jennings}.
These systems have been used to test Lorentz invariance
to parts in $10^{11}$ in gravity \cite{shao_grav1,shao_grav2}
and matter-gravity couplings \cite{shao_matter}.
They've also been used to search for velocity-dependent
effects from dimension $d=5$ terms \cite{shao_d5}
and from $d=8$ cubic terms in the gravity sector of the SME \cite{shao_d8}.
We also note that Lorentz violation can be constrained
with non-binary pulsars \cite{pulsar1,pulsar2}.
Binary pulsar are unique among orbital tests
in that they provide clean access to neutron coefficients
for Lorentz violation
and are therefore complementary to tests involving
ordinary matter.

The perturbations driven at higher frequencies with $m>1$ are
\begin{eqnarray}
\de x_\rh &=&
-\frac{2R}{m^2(m^2-1)}
\re\Big(\big(m^2C^{[m]}_{\rh\rh} -2imC^{[m]}_{\vp\rh}\big)
e^{im\om t}\Big)  \ ,
\notag \\
\de x_\vp &=&
-\frac{2R}{m^2(m^2-1)}
\re\Big(\big((3+m^2)C^{[m]}_{\vp\rh} + 2imC^{[m]}_{\rh\rh}\big)
e^{im\om t}\Big)  \ ,
\notag \\
\de x_z &=&
-\frac{2R}{m^2-1} \re\Big(C^{[m]}_{z\rh}e^{im\om t}\Big) \ .
\end{eqnarray}
Unlike the $m=1$ case,
which gave a secular evolution of the orbit,
violations with $m\geq 2$ produce periodic deviations from
the conventional orbit.
For example, $\de x_z$ produces periodic oscillations
about of the average orbital plane.
Among the effects from $\de x_\rh$ and $\de x_\vp$ displacements
is a time-dependent change in the areal velocity
$\de\dot A = R\om \de x_\rh + \tfrac12 R\de\dot x_\vp
=R^2\om m^{-1} \im\big(C^{[m]}_{\vp\rh}e^{im\om t}\big)$,
violating Kepler's second law.
Again, these periodic variations are similar
to ones arising in the gravity sector of the SME \cite{qbk06}
and could be sought in planetary motion
or in lunar laser ranging.
Note that since $k\geq j \geq |m|$,
the effects of Lorentz violation
at frequencies greater than $2\om$
necessarily involve the orbital speed of the satellite.
Consequently,  unsuppressed speed-independent
periodic variations only arise at twice
the orbital frequency.

\subsection{Acoustic Resonators}
\label{RES}

This section considers the effects of
Lorentz violation in continuous media
with particular focus on 
acoustic resonances in piezoelectric materials.
For continuous media,
the Lorentz-violating hamiltonian density can be taken as
\begin{equation}
\de\H = -\sum_{kjm} \rh^{1-k} |\Pvec|^k
\syjm{}{jm}(\Phat)\, \cT{kjm}\ ,
\label{dh}
\end{equation}
where $\rh$ is the mass density of the material,
$\Pvec$ is the momentum density,
and $\Phat = \Pvec/|\Pvec|$.
Using the above,
we can find  modifications
to the Hamilton equations of motion.
Alternatively,
we could instead employ a lagrangian approach,
where the leading-order change
to the Lagrange density can be taken as
\begin{equation}
\de\cl = \sum_{kjm} \rh |\vvec|^k
\syjm{}{jm}(\vhat)\, \cT{kjm}\ ,
\label{dL}
\end{equation}
where $\vvec = \vvec(\xvec,t)$ is the local velocity of the medium,
and $\vhat = \vvec/|\vvec|$.
Denoting the mechanical displacement of the medium
at equilibrium position $\xvec$
in a body-fixed frame as $\uvec = \uvec(\xvec,t)$,
the velocity is then given by $\vvec = \dot\uvec$.
Note that $\vvec \approx \Pvec/\rh$,
but differs slightly from the usual result due to Lorentz violation.

The conventional Lagrange density
for a piezoelectric material is given by
\begin{equation}
\cl = \tfrac12 \rh \dot{\uvec}^2
- \tfrac12\S^{abcd} u^a_{,b} u^c_{,d}
+\tfrac12 \ep^{ab} \ph_{,a} \ph_{,b}
- e^{abc} \ph_{,a} u^b_{,c} \ ,
\end{equation}
where $u^a_{,b} = \prt u^a/\prt x^b$ are
spatial derivatives of the displacements $u^a$,
$\ph_{,a} = \prt\ph/\prt x^a$
is the gradient of the electric potential $\ph$,
$\S^{abcd}$ is the stiffness tensor,
$\ep^{ab}$ is the permittivity tensor,
and $e^{abc}$ is the piezoelectric tensor.
The stiffness tensor $\S^{abcd}$ is taken to be symmetric
in the first pair of indices and the last pair of indices
and symmetric under interchange of the pairs,
giving twenty-one independent components.
The permittivity tensor $\ep^{ab}$ is symmetric,
and the piezoelectric tensor $e^{abc}$ is symmetric
in the last two indices.
The equations of motion for the system
including Lorentz violation are given by
\begin{equation}
\rh \ddot u^a + \rh C^{ab} \ddot u^b = T^{ab}_{,b}\ ,
\quad
0 = D^a_{,a}\ ,
\label{eom}
\end{equation}
where
$T^{ab} = \S^{abcd}u^c_{,d} + e^{cab} \ph_{,c}$
is the stress tensor,
$D^a = -\ep^{ab}\ph_{,b} + e^{abc}u^b_{,c}$
is the electric displacement field,
and $C^{ab}$ is the Lorentz-violating tensor
from \rf{C} evaluated at velocity $\vvec = \dot{\uvec}$.

Periodic solutions to \rf{eom}
can be found using methods similar to those used for orbits.
Solutions with period $2\pi/\om$
will in general includes various harmonics
of the fundamental frequency $\om$.
To find them,
first expand each variable in Fourier modes:
$\uvec = \sum_m \uvec^{[m]} e^{im\om t}$,
$\ph = \sum_m \ph^{[m]} e^{im\om t}$,
and
$C^{ab} = \sum_m C^{[m]ab} e^{im\om t}$.
The equations of motion \rf{eom}
lead to a set of coupled equations relating
the various Fourier components,
which can be solved perturbatively.
However,
we are primarily interested in changes to the frequency $\om$,
which can be found using a simpler method.

We begin by assuming the solution with Lorentz violation $\uvec$ 
has frequency and amplitudes that are close
to those for a conventional solution $\uvec_0$.
Manipulating the equations of motion
for $\uvec$ and $\uvec_0$,
one can show the relation
\begin{equation}
\rh \int_V d^3x\,\big(
\ddot{\uvec} \cdot \uvec_0
+ \ddot{\uvec} \cdot C \cdot \uvec_0
- \uvec \cdot \ddot{\uvec}_0
\big)
= \int_{\prt V} d\sivec\cdot\big(
T\cdot\uvec_0
- T_0\cdot\uvec
- \Dvec_0 \ph
+ \Dvec \ph_0
\big) \ ,
\end{equation}
where the left-hand side is integrated
over the volume $V$ of the resonator,
and the right-hand side is integrated
over the surface $\prt V$.
The conventional stress tensor $T_0$ 
depends on $\uvec_0$, the conventional potential $\ph_0$,
and the conventional displacement field $\Dvec_0$.
We then assume the surface terms vanish,
giving
\begin{equation}
\int_V d^3x\, \big(
\ddot\uvec \cdot \uvec_0
- \uvec \cdot \ddot \uvec_0
\big)
= -\int_V d^3x\,
\ddot\uvec \cdot C \cdot \uvec_0 \ .
\label{res1}
\end{equation}
Assuming simple harmonic conventional solutions,
this expression oscillates
at frequencies $m\om \pm\om_0 = (m\pm1)\om_0+m\de\om$,
where $\de\om=\om-\om_0$ is the shift
in the fundamental frequency
from the usual frequency $\om_0$.
Writing the amplitudes as
$\uvec^{[m]} = \uvec_0^{[m]} + \de \uvec^{[m]}$,
where $\de\uvec^{[m]}$ is the change due to Lorentz violation,
we can expand the frequency components of \rf{res1}
in small parameters depending on coefficients for Lorentz violation.
The zeroth-order equations are identically satisfied.
The first-order equations give
\begin{equation}
\int_V d^3x\, \big(
(1-m^2)\de \uvec^{[m]} \cdot \uvec_0^{[\pm 1]}
-2m^2\frac{\de\om}{\om_0}\,\uvec_0^{[m]} \cdot \uvec_0^{[\pm 1]} \big)
= \int_V d^3x\,  
\sum_{m'} {m'}^2
\uvec_0^{[m']}\cdot
C^{[m-m']}
\cdot \uvec_0^{[\pm1]}\ .
\end{equation}
Note that $\uvec_0^{[m]}=0$ for $m\neq \pm1$
since we assume $\uvec_0$ is simple harmonic.
The shift in frequency $\de\om$
can be isolated by taking $m=\mp1$,
which gives
\begin{eqnarray}
\frac{\de\om}{\om_0}
&\approx& -\frac12 \frac{\int_V d^3x\, \big(
\uvec_0^{[-1]}\cdot C^{[0]}\cdot \uvec_0^{[1]} +
\uvec_0^{[1]}\cdot C^{[-2]}\cdot \uvec_0^{[1]}
\big)}{
\int_V d^3x\,\uvec_0^{[1]} \cdot \uvec_0^{[-1]}}
\notag \\
&=& -\frac12
\frac{\int_V d^3x\, \vev{\uvec_0\cdot C\cdot \uvec_0}_t}{\int_V d^3x\, \vev{\uvec_0\cdot \uvec_0}_t} \ ,
\label{dw1}
\end{eqnarray}
where brackets $\vev{}_t$ indicate the time average.
The $C$ tensor in this expression is calculated using
the conventional velocity $\vvec_0 = \dot\uvec_0$.
The leading-order frequency shift is then completely determined
by the coefficients for Lorentz violation
and the usual solution $\uvec_0$.

The time averages in \rf{dw1}
may be difficult to calculate in general
but are relatively simple in the case
of standing waves with local linear polarization,
where we can take
$\uvec_0(\xvec,t) \rightarrow \uvec_0(\xvec)\sin(\om_0 t)$.
The velocity is replaced with
$\vvec_0(\xvec,t) \rightarrow \om_0\uvec(\xvec)\cos(\om_0 t)$
and is parallel to the displacement $\uvec_0(\xvec)$.
The time average in the denominator of \rf{dw1} becomes
$\vev{\uvec_0\cdot\uvec_0}_t \rightarrow \tfrac12 \uvec_0^2$.
The numerator can be shown to vanish for odd values of $k$.
For fixed even values of $k$,
the time average in the numerator becomes
\begin{eqnarray}
\vev{\uvec_0\cdot C\cdot \uvec_0}_t
&\rightarrow&
\vev{\sin^2(\om_0 t)\cos^{k-2}(\om_0 t)}_t\,
\uvec_0\cdot C(\om_0\uvec_0)\cdot\uvec_0
\notag \\
&=&
\frac{(k-3)!!}{k!!}
\uvec_0\cdot C(\om_0\uvec_0)\cdot\uvec_0
\end{eqnarray}
The frequency shift is then given by
\begin{equation}
\frac{\de\om}{\om_0} \approx
-\sum_{kjm} 
\om_0^{k-2}\frac{(k-1)!!}{(k-2)!!}
\frac{\int_V d^3x\,|\uvec_0|^k\yjm{jm}(\uhat_0)}{\int_V d^3x\, |\uvec_0|^2}
\cTlab{kjm}
\label{dw2}
\end{equation}
where $k$ is restricted to even values $k\geq2$,
and $\cTlab{kjm}$ are laboratory-frame coefficients.
The dimensionless factors multiplying the $\cT{kjm}$ coefficients
determine the sensitivity of an acoustic-resonator experiment.
Assuming oscillation amplitudes on the order of 100 angstrom
and frequencies on the order of a MHz,
these factors scale as $\sim 10^{-10(k-2)}$.
This drastically reduces the sensitivity
to violations with $k\neq 2$.
We therefore focus on the $k=2$ case.
The problem simplifies even further for cases
in which the vibration direction
$\uhat_0(\xvec)$ is relatively constant
over the volume of the resonator:
\begin{eqnarray}
\frac{\de\om}{\om_0} &\approx&
-\sum_{jm} \yjm{jm}(\uhat_0)\, \cTlab{2jm}
\notag \\
&\approx&
\half \sum_{dljm} M_n^{d-4}
\bc{(d+2l-5)/2}{l} 
\yjm{jm}(\uhat_0)\,
\big(\attlab{d}{(2-2l)jm} - \cttlab{d}{(2-2l)jm}\big)\ ,
\label{dw3}
\end{eqnarray}
assuming in the last line that the medium
is comprised of roughly equal
numbers of neutrons, protons, and electrons.
The frequency shift is then limited
to quadrupole $j=2$  and isotropic $j=0$ violations.

The rotational and orbital motion of the Earth
implies that the laboratory frame is noninertial.
As a result, the laboratory-frame coefficients
will change as the orientation and velocity
of the laboratory changes,
producing periodic variations in the frequency shift.
We account for these changes
using the transformation between
the laboratory frame and Sun-centered
frame discussed in Section \ref{LT}.
The rotations introduce sidereal variations
in the laboratory-frame coefficients.
The coefficients also vary with the angle of
the resonator in the laboratory.
In experiments involving rotating turntables,
this produces variations at the turn rate.
Annual changes in the velocity of Earth
lead to annual variations in the signal.
These, however, enter through boosts
and are suppressed by the boost velocity
$\be \approx 10^{-4}$
relative to the other variations.

The fluctuations in the frequency shift take the form
\begin{equation}
\frac{\de\om}{\om_0} = \sum_{\mt\ms\ma} A_{\mt\ms\ma}
e^{i\mt\vp + i\ms\om_\oplus T_\oplus + i \ma\Om_\oplus T} \ ,
\label{dw4}
\end{equation}
where
$\om_\oplus = 2\pi/\text{23hr 56min}$ and
$\Om_\oplus = 2\pi/\text{1yr}$
are respectively the sidereal and annual frequencies.
The time $T_\oplus$ is defined so that the laboratory zenith
points at right ascension $\al=0$ when $T_\oplus=0$,
and time $T=0$ at the vernal equinox.
The angle $\vp$ is between
the laboratory-frame $x$ axis and south.
The laboratory frame is fixed to the resonator,
which may be affixed to a turntable.
In this case,
$\vp$ changes at the turntable rotation rate $\omt$.
The indices $\mt$, $\ms$, and $\ma$
are the harmonic numbers for variations
at respectively
the turntable rotation frequency $\omt$,
sidereal frequency $\om_\oplus$,
and annual frequency $\Om_\oplus$.
The amplitudes obey the relation
$A^*_{\mt\ms\ma} = A_{(-\mt)(-\ms)(-\ma)}$,
ensuring that the frequency shift is real.

Applying the Lorentz transformations
outlined in Section \ref{LT}
to the laboratory-frame coefficients in \rf{dw3},
we find that the modulation amplitudes
due to rotations only are
\begin{equation}
A_{\mt\ms(\ma=0)} =
\half\sum_{dlj}
M_n^{d-4}
\bc{(d+2l-5)/2}{l}
\yjm{j\mt}(\uhat_0)\,
d^{(j)}_{\mt\ms}(-\ch)\,
\Big(\att{d}{(2-2l)j\ms}-\ctt{d}{(2-2l)j\ms}\Big) \ ,
\end{equation}
in terms of the Sun-frame
$\att{d}{kjm}$ and  $\ctt{d}{kjm}$ coefficients.
The isotropic $j=0$ violations produce a constant shift.
The quadrupole $j=2$ violations
give variations at frequencies
$\mt\omt + \ms\om_\oplus$
up to the second harmonic 
in both the turntable and sidereal rates:
$|\mt|,|\ms|\leq 2$.
Note, however,
that $|\mt|=1$ variations will be absent
in oscillators with horizontal or vertical vibrations.

Including leading-order boost effects due to
the orbital motion of the Earth gives
variations at frequencies 
$\mt\omt + \ms\om_\oplus + \ma\Om_\oplus$
with $\ma=\pm1$.
The amplitudes for these are given by
\begin{eqnarray}
A_{\mt\ms(\ma=\pm1)} &=&
\half \sum_{dlj}
M_n^{d-4}
\bc{(d+2l-5)/2}{l}
\yjm{j\mt}(\uhat_0)\,
d^{(j)}_{\mt\ms}(-\ch)\,
\notag \\
&& \qquad \times
\sum_{k'j'm'm''}
\G{d}{(2-2l)j\ms}{k'j'm'}{m''}\,
B_{m''\ma}
\Big(\att{d}{k'j'm'}-\ctt{d}{k'j'm'}\Big) \ ,
\end{eqnarray}
where the numerical $\G{d}{(2-2l)j\ms}{k'j'm'}{m''}$
constants are given in \rf{Gamma},
and the $B_{m''\ma}$ boost factors are in \rf{Bs}.
This gives sensitivity to other coefficients for Lorentz violation
but at levels suppressed by the small boost velocity
$\be\approx 10^{-4}$ of the Earth.

Searches for Lorentz violation in quartz resonators
have demonstrated sensitivities on the order of parts in $10^{14}$
to $d=4$ violations \cite{quartz1}
and are expected to improve by two orders of magnitude \cite{quartz2}.
We therefore expect sensitivities near $10^{-16}\text{ GeV}^{4-d}$
to the dimension-$d$ combinations $\att{d}{2jm}$ and $\ctt{d}{2jm}$.

\section{Summary}
\label{SUMMARY}

A violation of Lorentz invariance
would necessarily indicate new physics
with potential origins in quantum gravity.
High-precision experiments have limited
violations in a large variety of systems \cite{tables}.
In this paper,
we derive the effects of Lorentz violation
on dynamics of ordinary matter.
We include all linear dimension-$d$ violations in
the electrons, protons, and neutrons,
excluding violations involving
electromagnetic and gravitational interactions.

The effective hamiltonian for a macroscopic test body
is derived in Section \ref{HAMILTONIAN}.
The Lorentz-violating contributions are given
in \rf{dH} in terms of macroscopic coefficients
for Lorentz violation $\cT{kjm}$.
Equation \rf{cT2} relates these coefficients
to underlying SME coefficients
for electrons, protons, and neutrons.
Ignoring internal kinetic energy,
the result reduces to \rf{cT3}.
Equation \rf{cT4}
gives $\cT{kjm}$ for matter with
equal numbers electrons and protons,
and \rf{cT5} is for matter with equal
numbers of electrons, protons, and neutrons.
The equations of motion
are discussed in Section \ref{EOM}.
A modified Newton's second law
is given in \rf{N2}.
Section \ref{LT} discusses observer Lorentz
transformations of the coefficients,
relating coefficients in the Sun-centered
equatorial frame to a standard laboratory frame.
The boosts are calculated to first order in velocity,
resulting in \rf{lt}.

Section \ref{APPLICATIONS} contains several applications.
Tests of the the weak equivalence principle are
discussed in Section \ref{WEP},
including tests involving
free-fall experiments
\cite{ff87,ff89,ff90,ff92,ff96},
the space-based MICROSCOPE experiment
\cite{microscope}
and torsion-balance experiments
\cite{Roll,Braginskii,Adelberger,Su,Schlamminger}.
Implied bounds on isotropic
Lorentz violation from these experiments
are given in Table \ref{eptable},
demonstrating sensitivities
down to $\sim 10^{-13}\text{ GeV}^{4-d}$
to dimension $d$ violations.

Planetary orbits are
discussed in Section \ref{ORBITS}.
The effects of Lorentz violation
include a drift in eccentricity,
a rotation of the orbital plane,
and periodic variations about conventional orbits.
The small eccentricities of Earth and Venus
limit Lorentz violation at the $\sim 10^{-9}\text{ GeV}^{4-d}$ level.
The approximate alignment of the planets' orbital planes
leads to bounds of $\sim 10^{-12}\text{ GeV}^{4-d}$.
Improvements on these rough constraints
are expected in detailed studies of
planetary ephemerides \cite{iorio,hees}
and through lunar laser ranging \cite{llr1,llr2,llr3}.
Binary pulsars offer another promising
area of study that is particularly
sensitive to Lorentz violations in neutrons
\cite{qbk06,jennings,shao_grav1,shao_grav2,shao_matter,shao_d5,shao_d8}.

Section \ref{RES} gives the Lorentz-violating
Lagrange density for continuous media \rf{dL}.
The shift in resonant frequency in piezoelectric
acoustic resonators is calculated,
including boost effects.
The shifts vary periodically at frequencies
involving the 
the turntable rotation rate,
the Earth's sidereal rotation rate,
and the Earth's orbital frequency.
Experiments have demonstrated sensitivities
at parts in $10^{14}$ to dimension $d=4$ Lorentz violations \cite{quartz1}
and are expected to reach $10^{-16}\text{ GeV}^{4-d}$
to arbitrary dimension $d$ violations \cite{quartz2}.

These results show that extreme precision can be achieve
in studies of spacetime symmetries in macroscopic matter.
While not as sensitive as the best of the microscopic tests
\cite{atomse1,atomse2,atomse3,atomspn1,atomspn2},
experiments involving ordinary matter rely on different assumptions
and may provide access to different combinations of SME coefficients
and therefore represent a powerful tool in our search for new physics.

\funding{
  This research was funded by
  the United States National Science Foundation
  grant number PHY-1819412.
}

\conflictsofinterest{The author declares no conflicts of interest.}

\appendixtitles{no}
\appendix
\section{}
\label{appendix}

This Appendix derives the symmetric
product identity for spherical-harmonic tensors.
See Ref.\ \cite{Yrjm}
for a detailed discussion of the $\Yrjm{\vr}{jm}$ tensors
and the notation used here.

Expanding the symmetric product of two spherical-harmonic tensors
in the basis of spherical-harmonic tensors,
we can write
\begin{equation}
\Yrjm{\vr_1}{j_1m_1}\odot\Yrjm{\vr_2}{j_2m_2}
= \sum_{JM}\A{\vr_1\vr_2}{j_1m_1j_2m_2JM}\Yrjm{(\vr_1+\vr_2)}{JM}\ .
\label{Yprod}
\end{equation}
The $\A{\vr_1\vr_2}{j_1m_1j_2m_2JM}$ coefficients 
are nonzero for the usual angular-momentum-addition relations
$M=m_1+m_2$ and $j_1+j_2\geq J \geq |j_1-j_2|$
and for $j_1+j_2-J = \text{even}$.
The nonzero values are real and given by
\begin{eqnarray}
\A{\vr_1\vr_2}{j_1m_1j_2m_2JM}
&=&\Yrjmconj{(\vr_1+\vr_2)}{JM} \cdot (\Yrjm{\vr_1}{j_1m_1} \odot\Yrjm{\vr_2}{j_2m_2})
\notag \\
&=&
(2J+1)(J+M)!(J-M)!
\frac{\B{\vr_1+\vr_2}{JM}}{\B{\vr_1}{j_1m_1}\B{\vr_2}{j_2m_2}}
\notag\\
&&\times
\tfrac{(j_1+j_2-J-1)!!(j_1-j_2+J-1)!!(j_2-j_1+J-1)!!}{(j_1+j_2+J+1)!!}
\notag\\
&&\times
\sum_n
\tfrac{(-1)^n}
{n!(n+J-j_2-m_1)!(n+J-j_1+m_2)!(j_1+j_2-J-n)!(j_1+m_1-n)!(j_2-m_2-n)!} \ ,
\notag\\
\label{Acoeffs}
\end{eqnarray}
where the sum is limited to $n$ values
that give nonnegative arguments in all factorials,
and we define
\begin{equation}
\B{\vr}{jm} =
(-1)^{j/2}
\sqrt{\tfrac{(\vr+j+1)!!(\vr-j)!!}{(2j+1)\vr!(j+m)!(j-m)!}}\ ,
\end{equation}
for convenience.

Our derivation starts by considering the vector
\begin{equation}
\mbf\xi =
\frac{i\ze}{\sqrt2}\Eu
+\frac{i}{\sqrt2\ze}\Ed
+\Ez 
\end{equation}
for arbitrary complex number $\ze$.
The unit vectors $\Eu = (\Ex+i\Ey)/\sqrt2$,
$\Ed=(\Ex-i\Ey)/\sqrt2$, and $\Ez$
form a spin-eigenbasis for quantization
along the $z$ axis.
A short calculation reveals
that the $j$-fold symmetric product
of $\mbf\xi$ is
\begin{equation}
\mbf\xi^{\,\odot j}
= \sum_m C_{jm} \ze^m \Yjm{jm} \ ,
\label{xij}
\end{equation}
where $\Yjm{jm} = \Yrjm{j}{jm}$
are the traceless spherical-harmonic tensors,
and
\begin{equation}
C_{jm} = (-i)^m \sqrt{\tfrac{j!(2j-1)!!}{(j+m)!(j-m)!}}
\end{equation}
for $|m|\leq j$.
The symmetric product in \rf{xij}
serves as a generating function
for the traceless spherical-harmonic tensors.
Note that $\mbf\xi\cdot\mbf\xi = 0$,
which confirming that it is traceless.

Next consider the inner product
\begin{equation}
\mbf\xi^{\,\odot J} \cdot(\Yjm{j_1m_1}^*\odot\Yjm{j_2m_2}^*)
=
(\mbf\xi^{\,\odot j_1}\cdot \Yjm{j_1m_1}^*)
(\mbf\xi^{\,\odot j_2}\cdot \Yjm{j_2m_2}^*) \ ,
\end{equation}
where $J=j_1+j_2$.
The two sides of this equation evaluate to
\begin{equation}
\sum_M C_{JM} \ze^M \Yjm{JM} \cdot(\Yjm{j_1m_1}^*\odot\Yjm{j_2m_2}^*)
=
C_{j_1m_1}C_{j_2m_2} \ze^{m_1+m_2} \ ,
\end{equation}
which implies
\begin{equation}
\Yjm{JM} \cdot(\Yjm{j_1m_1}^*\odot\Yjm{j_2m_2}^*)
= \de_{M,m_1+m_2}\frac{C_{j_1m_1}C_{j_2m_2}}{C_{JM}}  \ .
\label{ident1}
\end{equation}
The complex conjugate of this gives
the $\A{j_1j_2}{j_1m_1j_2m_2JM}$ coefficients
for traceless tensors.

To find the inner product for tensors of nonzero trace,
we consider traces of the product
\begin{equation}
\mbf\xi_1^{\odot j_1}\odot\mbf\xi_2^{\odot j_2}
= \sum_{m_1m_2} C_{j_1m_1}C_{j_1m_1} \ze_1^{m_1}\ze_2^{m_2}
\Yjm{j_1m_1}\odot\Yjm{j_2m_2}\ .
\end{equation}
Using
$\mbf\xi_1\cdot\mbf\xi_2 = -(\ze_1-\ze_2)^2/2\ze_1\ze_2$,
one can show that taking $N$ traces gives
\begin{equation}
(-\ze_1\ze_2)^N g^{\odot N}\cdot(\mbf\xi_1^{\odot j_1}\odot\mbf\xi_2^{\odot j_2})
=
\frac{j_1^{\underline N}j_2^{\underline N}}{(j_1+j_2)^{\underline{2N}}}
(\ze_1-\ze_2)^{2N}
\mbf\xi_1^{\odot(j_1-N)}\odot\mbf\xi_2^{\odot(j_2-N)}\ ,
\end{equation}
where $g$ is the euclidean metric,
and $x^{\underline n}$ indicates the falling factorial.
Matching terms by their powers in $\xi_1$ and $\xi_2$,
we find
\begin{eqnarray}
g^{\odot N}\cdot(\Yjm{j_1m_1}\odot\Yjm{j_2m_2}) &=&
\frac{(2N)!(-1)^Nj_1^{\underline N}j_2^{\underline N}}{(j_1+j_2)^{\underline{2N}}C_{j_1m_1}C_{j_2m_2}}
\sum_{n=0}^{2N}
\tfrac{(-1)^n}{n!(2N-n)!}C_{(j_1-N)(m_1+N-n)}C_{(j_2-N)(m_2-N+n)}
\notag \\
&&\times
\Yjm{(j_1-N)(m_1+N-n)}\odot\Yjm{(j_2-N)(m_2-N+n)}\ .
\label{ident2}
\end{eqnarray}

Finally, the identities
\begin{eqnarray}
\Yrjm{\vr}{jm} &=& \D{\vr}{j} \Yjm{jm} \odot g^{\odot \frac12(\vr-j)} \ ,
\notag \\
\D{\vr}{j}g^{\odot N}\cdot\Yrjm{\vr}{jm}
&=& \D{\vr-2N}{j}
\Yrjm{\vr-2N}{jm}\ , 
\end{eqnarray}
where
\begin{equation}
\D{\vr}{j} = \sqrt{\tfrac{\vr!(2j+1)!!}{j!(\vr+j+1)!!(\vr-j)!!}}\ ,
\end{equation}
can be used to show that
\begin{equation}
\D{\vr_1+\vr_2}{J}
\Yrjmconj{(\vr_1+\vr_2)}{JM} \cdot
(\Yrjm{\vr_1}{j_1m_1} \odot\Yrjm{\vr_2}{j_2m_2})
=
\D{\vr_1}{j_1}\D{\vr_2}{j_2}\D{j_1+j_2}{J}\D{j_1+j_2}{J}
\Yjm{JM}^*\cdot \big(g^{\odot\frac12(j_1+j_2-J)}
\cdot(\Yjm{j_1m_2}\odot\Yjm{j_2m_2})\big) \ .
\label{ident3}
\end{equation}
Combining \rf{ident3}
with identities \rf{ident2}
and  \rf{ident1}
yields the final result in
\rf{Yprod} and \rf{Acoeffs}.

\reftitle{References}


\begin{thebibliography}{999}


\bibitem{ks}
  Kosteleck\'y, V.A.; Samuel, S.
  Spontaneous Breaking of Lorentz Symmetry in String Theory.
  {\em Phys.\ Rev.\ D} {\bf 1989}, {\em 39}, 683.
\bibitem{kp}
  Kosteleck\'y, V.A.; Potting, R.
  CPT and strings.
  {\em Nucl.\ Phys.\ B} {\bf 1991}, {\em 359}, 545.

\bibitem{ck1}
  Colladay, D.; Kosteleck\'y, V.A.
  CPT violation and the standard model.
  {\em Phys.\ Rev.\ D} {\bf 1997}, {\em 55}, 6760.
\bibitem{ck2}
  Colladay, D.; Kosteleck\'y, V.A.
  Lorentz violating extension of the standard model.
  {\em Phys.\ Rev.\ D} {\bf 1998}, {\em 58}, 116002.
\bibitem{smegrav}
  Kosteleck\'y, V.A.
  Gravity, Lorentz violation, and the standard model.
  {\em Phys.\ Rev.\ D} {\bf 2004}, {\em 69}, 105009.
  
\bibitem{bluhm_rev}
  Bluhm, R.
  Overview of the SME: Implications and phenomenology of Lorentz violation.
  {\em Lect.\ Notes Phys.} {\bf 2006}, {\em 702}, 191.
\bibitem{tasson_rev}
  Tasson, J.D.
  What Do We Know About Lorentz Invariance?.
  {\em Rept.\ Prog.\ Phys.} {\bf 2014}, {\em 77}, 062901.
\bibitem{hees_rev}
  Hees, A.; Bailey, Q.G.; Bourgoin, A.; Bars, H.P.L.;
  Guerlin, C.; Le Poncin-Lafitte, C.
  Tests of Lorentz symmetry in the gravitational sector.
  {\em Universe} {\bf 2016}, {\em 2}, 30.

\bibitem{tables}
  Kosteleck\'y, V.A.; Russell, N.
  Data Tables for Lorentz and CPT Violation.
  {\em Rev.\ Mod.\ Phys.} {\bf 2011}, {\em 83}, 11.


\bibitem{rbk05}
  Bluhm, R.; Kosteleck\'y, V.A.
  Spontaneous Lorentz violation, Nambu-Goldstone modes, and gravity.
  {\em Phys.\ Rev.\ D} {\bf 2005}, {\em 71}, 065008.

\bibitem{rbk08}
  Bluhm, R.; Fung, W.H.; Kosteleck\'y, V.A.
  Spontaneous Lorentz and Diffeomorphism Violation, Massive Modes, and Gravity.
  {\em Phys.\ Rev.\ D} {\bf 2008}, {\em 77}, 065020.

\bibitem{qbk06}
  Bailey, Q.G.; Kosteleck\'y, V.A.
  Signals for Lorentz violation in post-Newtonian gravity.
  {\em Phys.\ Rev.\ D} {\bf 2006}, {\em 74}, 045001.

\bibitem{km09}
  Kosteleck\'y, V.A.; Mewes, M.
  Electrodynamics with Lorentz-violating operators of arbitrary dimension.
  {\em Phys.\ Rev.\ D} {\bf 2009}, {\em 80}, 015020.

\bibitem{km12}
  Kosteleck\'y, V.A.; Mewes, M.
  Neutrinos with Lorentz-violating operators of arbitrary dimension.
  {\em Phys.\ Rev.\ D} {\bf 2012}, {\em 85}, 096005.

\bibitem{km13}
  Kosteleck\'y, V.A.; Mewes, M.
  Fermions with Lorentz-violating operators of arbitrary dimension.
  {\em Phys.\ Rev.\ D} {\bf 2013}, {\em 88}, 096006.

\bibitem{km16}
  Kosteleck\'y, V.A.; Mewes, M.
  Testing local Lorentz invariance with gravitational waves.
  {\em Phys.\ Lett.\ B} {\bf 2016}, {\em 757}, 510.

\bibitem{kt09}
  Kosteleck\'y, V.A.;  Tasson, J.
  Prospects for Large Relativity Violations in Matter-Gravity Couplings.
  {\em Phys.\ Rev.\ Lett.} {\bf 2009}, {\em 102}, 010402.

\bibitem{ctrans1}
  Colladay, D.; McDonald, P.
  Redefining spinors in Lorentz violating QED.
  {\em  J.\ Math.\ Phys.} {\bf 2002}, {\em 43}, 3554.
\bibitem{ctrans2}
  Kosteleck\'y, V.A.; Mewes, M.
  Signals for Lorentz violation in electrodynamics.
  {\em Phys.\ Rev.\ D} {\bf 2002}, {\bf 66}, 056005.
  
\bibitem{bire1}
  Kosteleck\'y, V.A.; Mewes, M.
  Cosmological constraints on Lorentz violation in electrodynamics.
  {\em Phys.\ Rev.\ Lett.} {\bf 2001}, {\em 87}, 251304.
\bibitem{bire2}
  Kosteleck\'y, V.A.; Mewes, M.
  Sensitive polarimetric search for relativity violations in gamma-ray bursts.
  {Phys.\ Rev.\ Lett.} {\bf 2006}, {\em 97}, 140401.
\bibitem{bire3}
  Kosteleck\'y, V.A.; Mewes, M.
  Lorentz-violating electrodynamics and the cosmic microwave background.
  {Phys.\ Rev.\ Lett.} {\bf 2007}, {\em 99}, 011601.
\bibitem{bire4}
  Brown, M.L.; Ade, P.; Bock, J.; Bowden, M.; Cahill, G.; Castro, P.G.;
  Church, S.; Culverhouse, T.; Friedman, R.B.; Ganga, K.; et al.
  Improved measurements of the temperature and polarization of the CMB from QUaD.
  {\em Astrophys.\ J.} {\bf 2009}, {\em 705}, 978.
\bibitem{bire5}
  Hinshaw, G.; Larson, D.; Komatsu, E.; Spergel, D.N.; Bennett, C.L.;
  Dunkley, J.; Nolta, M.R.; Halpern, M.; Hill, R.S.; Odegard, N.; et al.
  Nine-Year Wilkinson Microwave Anisotropy Probe (WMAP) Observations:
  Cosmological Parameter Results.
  {\em Astrophys.\ J.\ Suppl.} {\bf 2013}, {\em 208}, 19.
\bibitem{bire6}
  Kosteleck\'y, V.A.; Mewes, M.
  Constraints on relativity violations from gamma-ray bursts.
  {Phys.\ Rev.\ Lett.} {\bf 2013}, {\em 110}, 201601.
\bibitem{bire7}
  Aghanim, N.; Ashdown, M.; Aumont, J.; Baccigalupi, C.; Ballardini, M.;
  Banday, A.J.; Barreiro, R.B.; Bartolo, N.; Basak, S.; Benabed, K.; et al.
  Planck intermediate results. XLIX.
  Parity-violation constraints from polarization data.
  {\em Astron.\ Astrophys.} {\bf 2016}, {\em 596}, A110.
\bibitem{bire8}
  Kislat, F.;
  Constraints on Lorentz Invariance Violation
  from Optical Polarimetry of Astrophysical Objects.
  {\em Symmetry} {\bf 2018}, {\em 10}, 596.
\bibitem{bire9}
  Friedman, A.S.; Leon, D.; Crowley, K.D.; Johnson, D.; Teply, G.;
  Tytler, D.; Keating, B.G.; Cole, G.M.
  Constraints on Lorentz Invariance and $CPT$ Violation
  using Optical Photometry and Polarimetry of Active Galaxies BL Lacertae
  and S5 B0716+714.
  {\em Phys.\ Rev.\ D} {\bf 2019}, {\em 99}, 035045.
\bibitem{bire10}
  Pogosian, L.; Shimon, M.; Mewes, M.; Keating, B.
  Future CMB constraints on cosmic birefringence
  and implications for fundamental physics.
  {\em Phys.\ Rev.\ D} {\bf 2019}, {\em 100}, 023507.
\bibitem{bire11}
  Friedman, A.S; Gerasimov, R.; Kislat, F.; Leon, D.;
  Stevens, W.; Tytler, E.; Keating, B.G.
  Improved Constraints on Anisotropic Birefringent Lorentz Invariance
  and CPT Violation from Broadband Optical Polarimetry of High Redshift Galaxies.
  {\bf Phys.\ Rev.\ D} {\bf 2020}, {\em 102}, 043008.
  
\bibitem{qed1}
  Ding, Y.; Kosteleck\'y, V.A.
  Lorentz-violating spinor electrodynamics and Penning traps.
  {\em Phys.\ Rev.\ D} {\bf 2016}, {\em 94}, 056008.
\bibitem{qed2}
  Kosteleck\'y, V.A.; Li, Z.
  Gauge field theories with Lorentz-violating operators of arbitrary dimension.
  {\em Phys.\ Rev.\ D}, {\bf 2019}, {\em 99}, 056016.

\bibitem{li20}
  Kosteleck\'y, V.A.; Li, Z.
  Backgrounds in gravitational effective field theory.
  arXiv:2008.12206.
\bibitem{kt11}
  Kosteleck\'y, V.A.;  Tasson, J.
  Matter-gravity couplings and Lorentz violation.
  {\em Phys.\ Rev.\ D}, {\bf 2011}, {\em 83}, 016013.

\bibitem{astro1}
  Altschul, B.
  Synchrotron and inverse compton constraints on Lorentz violations for electrons.
  {\em Phys.\ Rev.\ D} {\bf 2006}, {\em 74}, 083003.
\bibitem{astro2}
  Altschul, B.
  Astrophysical limits on Lorentz violation for all charged species.
  {\em Astropart.\ Phys.} {\bf 2007}, {\em 28}, 380.
\bibitem{astro3}
  Altschul, B.
  Limits on Neutron Lorentz Violation from the Stability
  of Primary Cosmic Ray Protons.
  {\em Phys.\ Rev.\ D} {\bf 2008}, {\em 78}, 085018.
\bibitem{astro4}
  Stecker, F.W.
  Limiting superluminal electron and neutrino velocities using
  the 2010 Crab Nebula flare and the IceCube PeV neutrino events.
  {\em Astropart.\ Phys.} {\bf 2014}, {\em 56}, 16.
\bibitem{astro5}
  Satunin, P.
  One-loop correction to the photon velocity in Lorentz-violating QED.
  {\em Phys.\ Rev.\ D} {\bf 2018}, {\em 97}, 125016.

\bibitem{shao_matter}
  Shao, L.
  Lorentz-Violating Matter-Gravity Couplings in Small-Eccentricity Binary Pulsars.
  {\em Symmetry} {\bf 2019}, {\em 11}, 1098.

\bibitem{pulsar1}
  Altschul, B.
  Limits on Neutron Lorentz Violation from Pulsar Timing.
  {\em Phys.\ Rev.\ D} {\bf 2007}, {\em 75}, 023001.
  
\bibitem{wep1}
  Hohensee, M.A.; Leefer, N.; Budker, D.; Harabati, C.;
  Dzuba V.A.; Flambaum, V.V.
  Limits on Violations of Lorentz Symmetry and the Einstein Equivalence Principle
  using Radio-Frequency Spectroscopy of Atomic Dysprosium.
  {\em Phys.\ Rev.\ Lett.} {\bf 2013}, {\em 111}, 050401.
\bibitem{wep2}
  Hohensee, M.A.; Mueller, H.; Wiringa, R.B.
  Equivalence Principle and Bound Kinetic Energy.
  {\em Phys.\ Rev.\ Lett.} {\bf 2013}, {\em 111}, 151102.
  
\bibitem{gravimeters}
  Flowers, N.A.; Goodge, C.; Tasson, J.D.
  Superconducting-Gravimeter Tests of Local Lorentz Invariance.
  {\em Phys.\ Rev.\ Lett.} {\bf 2017}, {\em 119}, 201101.
  
\bibitem{accel1}
  Lane, C.D.
  Probing Lorentz violation with Doppler-shift experiment.
  {Phys.\ Rev.\ D} {\bf 2005}, {\em 72}, 016005.
\bibitem{accel2}
  Altschul, B.
  Laboratory Bounds on Electron Lorentz Violation.
  {Phys.\ Rev.\ D} {\bf 2010}, {\em 82}, 016002.
\bibitem{accel3}
  Botermann, B.; Bing, D.; Geppert, C.; Gwinner, G.; H\"ansch, T.W.;
  Huber, G.; Karpuk, S.; Krieger, A.; K\"uhl, T.; N\"ortersh\"auser, W.; et al.
  Test of Time Dilation Using Stored Li$^+$ Ions as Clocks at Relativistic Speed.
  {\em Phys.\ Rev.\ Lett.} {\bf 2014}, {\bf 113}, 120405.

\bibitem{cavities1}
  Muller, H.; Herrmann, S.; Saenz, A.; Peters, A.; Lammerzahl, C.
  Optical cavity tests of Lorentz invariance for the electron.
  {\em Phys.\ Rev.\ D} {\bf 2003}, {\em 68}, 116006.
\bibitem{cavities2}
  Muller, H.
  Testing Lorentz invariance by use of vacuum and matter filled cavity resonators.
  {\em Phys.\ Rev.\ D} {\bf 2005}, {\em 71}, 045004.
\bibitem{cavities3}
  Muller, H.; Stanwix, P.L.; Tobar, M.E.; Ivanov, E.; Wolf, P.;
  Herrmann, S.; Senger, A.; Kovalchuk, E.; Peters, A.
  Relativity tests by complementary rotating Michelson-Morley experiments.
  {Phys.\ Rev.\ Lett.} {\bf 2007}, {\em 99}, 050401.

\bibitem{atoms1}
  Kosteleck\'y, V.A.; Lane, C.D.
  Constraints on Lorentz violation from clock comparison experiments.
  {\em Phys.\ Rev.\ D} {\bf 1999}, {\em 60}, 116010.
\bibitem{atoms2}
  Wolf, P.; Chapelet, F.; Bize, S.; Clairon, A.
  Cold Atom Clock Test of Lorentz Invariance in the Matter Sector.
  {Phys.\ Rev.\ Lett.} {\bf 2006}, {\em 96}, 060801.
\bibitem{atoms3}
  Altschul, B.
  Testing Electron Boost Invariance with 2S-1S Hydrogen Spectroscopy.
  {\em Phys.\ Rev.\ D} {\bf 2010}, {\em 81}, 041701.
\bibitem{atoms4}
  Hohensee, M.A.; Chu, S.; Peters, A.; Muller, H.
  Equivalence Principle and Gravitational Redshift.
  {Phys.\ Rev.\ Lett.} {\bf 2011}, {\em 106}, 151102.
\bibitem{atoms5}
  Matveev, A. Parthey, C.G.; Predehl, K.; Alnis, J.; Beyer, A.;
  Holzwarth, R.; Udem, T.; Wilken, T.; Kolachevsky, N.; Abgrall, M.; et al.
  Precision Measurement of the Hydrogen 1S-2S Frequency via a 920-km Fiber Link.
  {\em Phys.\ Rev.\ Lett.} {\bf 2013}, {\em 110}, 230801.
\bibitem{atoms6}
 Dzuba, V.A.; Flambaum, V.V.
 Limits on gravitational Einstein equivalence principle violation
 from monitoring atomic clock frequencies during a year.
 {\em Phys.\ Rev.\ D} {\bf 2017}, {\em 95}, 015019.
\bibitem{atoms7}
  Pihan-Le Bars, H.; Guerlin, C.; Lasseri, R.D.; Ebran, J.P.; Bailey, Q.G.;
  Bize, S.; Khan, E.; Wolf, P.
  Lorentz-symmetry test at Planck-scale suppression
  with nucleons in a spin-polarized $^{133}$Cs cold atom clock.
  {\em Phys.\ Rev.\ D} {\bf 2017}, {\em 95}, 075026.

\bibitem{atomse1}
  Sanner, C.; Huntemann, N.; Lange, R.; Tamm, C.; Peik, E.;
  Safronova, M.S.; Porsev, S.G.
  Optical clock comparison for Lorentz symmetry testing.
  {\em Nature} {\bf 2019}, {\bf 567}, 204.
  
\bibitem{atomse2}
   Pruttivarasin, T.; Ramm, M.; Porsev, S.G.; Tupitsyn, I.I.; Safronova, M.;
  Hohensee, M.A.;  Haeffner, H.
  A Michelson-Morley Test of Lorentz Symmetry for Electrons.
  {\em Nature} {\bf 2015}, {\em 517}, 592.
\bibitem{atomse3}
  Megidish, E.; Broz, J.; Greene, N.; H\"affner, H.
  Improved Test of Local Lorentz Invariance from
  a Deterministic Preparation of Entangled States.
  {\em Phys.\ Rev.\ Lett.} {\bf 2019}, {\em 122}, 123605.

\bibitem{atomspn1}
  Smiciklas, M.; Brown, J.M.; Cheuk, L.W.; Romalis, M.V.
  A new test of local Lorentz invariance using $^{21}$Ne-Rb-K comagnetometer.
  {\em Phys.\ Rev.\ Lett.} {\bf 2011}, {\em 107}, 171604.
\bibitem{atomspn2}
  Flambaum, V.V.; Romalis, M.V.
  Effects of the Lorentz invariance violation
  on Coulomb interaction in nuclei and atoms.
  {\em Phys.\ Rev.\ Lett.} {\bf 2017}, {\em 118}, 142501.

\bibitem{quartz1}
  Lo, A.; Haslinger, P.; Mizrachi, E.; Anderegg, L.; M\:uller, H.;
  Hohensee, M.; Goryachev, M; Tobar, M.E.
  Acoustic tests of Lorentz symmetry using quartz oscillators.
  {\em Phys.\ Rev.\ X} {\bf 2016}, {\em 6}, 011018.

\bibitem{kv15}
  Kosteleck\'y, V.A.; Vargas, A.J.
  Lorentz and CPT tests with hydrogen, antihydrogen, and related systems.
  {\em Phys.\ Rev.\ D} {\bf 2015}, {\em 92}, 056002.
\bibitem{kv18}
  Kosteleck\'y, V.A.; Vargas, A.J.
  Lorentz and CPT Tests with Clock-Comparison Experiments.
  {\em Phys.\ Rev.\ D} {\bf 2018}, {\em 98}, 036003.
    
\bibitem{schreck16}
  Schreck, M.
  Classical Lagrangians and Finsler structures for
  the nonminimal fermion sector of the Standard-Model Extension.
  {\em Phys.\ Rev.\ D} {\bf 2016}, {\em 93}, 105017.
  
\bibitem{ding20}
  Ding, Y.; Rawnak, M.F.
  Lorentz and CPT tests with charge-to-mass ratio comparisons in Penning traps.
  {\em Phys.\ Rev.\ D} {\bf 2020}, {\em 102}, 056009.

\bibitem{Yrjm}
  Ledesma, F.G.; Mewes, M.
  Spherical-harmonic tensors.
  {\em Phys.\ Rev.\ Research} {\bf 2020}, {\em 2}, 043061.

\bibitem{bertschinger}
  Bertschinger, T.H.; Flowers, N.A.; Moseley, S.; Pfeifer, C.R.;
  Tasson, J.D.; Yang, S.
  Spacetime Symmetries and Classical Mechanics.
  {\em Symmetry} {\bf 2018}, {\em 11}, 22. 

\bibitem{clyburn}
  Clyburn, M.; Lane, C.D.
  Lorentz Violation at the Level of Undergraduate Classical Mechanics.
  {\em Symmetry} {\bf 2020}, {\em 12}, 1734.

\bibitem{ff86}
  Cavasinni, V.; Iacopini, E.; Polacco, E.; Stefanini, G.
  Galileo's experiment on free falling bodies using modern optical techniques.
  {\em Phys.\ Lett.\ A} {\bf 1986}, {\em 116}, 157.
  
\bibitem{ff87}
 Niebauer, T.M.; Mchugh, M.P.; Faller, J.E.
 Galilean Test for the Fifth Force.
 {\em Phys.\ Rev.\ Lett.} {\bf 1987}, {\em 59}, 609.

\bibitem{ff89}
  Kuroda, K.; Mio, N.
  Test of a composition-dependent force by a free-fall interferometer.
  {\em Phys.\ Rev.\ Lett.} {\bf 1989}, {\em 62}, 1941.
\bibitem{ff90}
  Kuroda, K.; Mio, N.
  Limits on a possible composition-dependent force by a Galilean experiment.
  {\em Phys.\ Rev.\ D} {\bf 1990}, {\em 42}, 3903.

\bibitem{ff92}
  Carusotto, S.; Cavasinni, V.; Mordacci, A.; Perrone, F.;
  Polacco, E.; Iacopini, E.; Stefanini, G.
  Test of the g universality with a Galileo's type experiment.
  {\em Phys.\ Rev.\ Lett.} {\bf 1992}, {\em 69}, 1722.
\bibitem{ff96}
  Carusotto, S.; Cavasinni, V.; Perrone, F.; Polacco, E.;
  Iacopini, E.; Stefanini, G.
  g-universality test with a Galileo's type experiment.
  {\em Nuovo Cim.\ B} {\bf 1996}, {\em 111}, 1259.

\bibitem{step}
  Overduin, J.; Everitt, F.; Worden, P.; Mester, J.
  STEP and fundamental physics.
  {\em Class.\ Quant.\ Grav.} {\bf 2012}, {\em 29}, 184012.
  
\bibitem{gg}
  Nobili, A.M; Shao, M.; Pegna, R.; Zavattini, G.; Turyshev, S.G.;
  Lucchesi, D.M.; De Michele, A.; Doravari, S.; Comandi, G.L; Saravanan, T.R.; et al.
  'Galileo Galilei' (GG): Space test of the weak equivalence principle
  to 10(-17) and laboratory demonstrations.
  {\em Class.\ Quant.\ Grav} {\bf 2012}, {\em 29}, 184011.

\bibitem{microscope}
  Touboul, P.; M\'etris, G.; Rodrigues, M.; Andr\'e, Y.; Baghi, Q.;
  Berg\'e, J.; Boulanger, D.; Bremer, S.; Carle, P.; Chhun, R.; et al.
  MICROSCOPE Mission: First Results of a Space Test of the Equivalence Principle.
  {\em Phys.\ Rev.\ Lett.} {\bf 2017}, {\em 119}, 231101.
  
\bibitem{eotvos}
  Von E\"otv\"os, R.
  \"Uber die anziehung der erde auf verschiedene substanzen.
  {\em Math.\ Naturwiss.\ Ber.\ Ung.} {\bf 1890}, {\em 8}, S65.

\bibitem{Roll}
  Roll, P.G.; Krotkov, R.; Dicke, R.H.
  The equivalence of inertial and passive gravitational mass.
  {\em Annals Phys.} {\bf 1964}, {\em 26}, 442.

\bibitem{Braginskii} 
  Braginskii, V.B; Panov, V.I.
  Verification of the equivalence of inertial and gravitational masses.
  {\em Sov.\ Phys.\ JETP} {\bf 1972}, {\em 34}, 463.
  
\bibitem{Adelberger}
  Adelberger, E.G.; Stubbs, C.W.; Heckel, B.R.; Su, Y.; Swanson, H.E.;
  Smith, G.; Gundlach, J.H.; Rogers, W.F.
  Testing the equivalence principle in the field of the Earth:
  Particle physics at masses beloved 1 $\mu$eV.
  {\em Phys.\ Rev.\ D} {\bf 1990}, {\em 42}, 3267.

\bibitem{Su}
  Su, Y.; Heckel, B.R.; Adelberger, E.G.; Gundlach, J.H.;
  Harris, M.; Smith, G.L.; Swanson, H.E.
  New tests of the universality of free fall.
  {\em Phys.\ Rev.\ D} {\bf 1994}, {\em 50}, 3614.
  
\bibitem{Schlamminger}
  Schlamminger, S.; Choi, K.Y.; Wagner, T.A.; Gundlach, J.H.; Adelberger, E.G.
  Test of the equivalence principle using a rotating torsion balance.
  {\em Phys.\ Rev.\ Lett.} {\bf 2008}, {\em 100}, 041101.

\bibitem{microscopeLV}
  Pihan-Le Bars, G.; Guerlin, C.; Hees, A.; Peaucelle, R.; Tasson, J.D.;
  Bailey, Q.G.; Mo, G.; Delva, P.; Meynadier, F.; Touboul, P.; et al.
  New Test of Lorentz Invariance Using the MICROSCOPE Space Mission.
  {\em Phys.\ Rev.\ Lett.} {\bf 2019}, {\em 123}, 231102.
  
\bibitem{ffdt}
  Sondag, A.; Dittus, H.
  Electrostatic Positioning System for a free fall test at drop tower
  Bremen and an overview of tests for the Weak Equivalence Principle
  in past, present and future.
  {\em Adv.\ Space Res.} {\bf 2016}, {\em 58}, 644.

\bibitem{ffballoon}
  Iafolla, V.; Nozzoli, S.; Lorenzini, E.C.; Milyukov, V.
  Methodology and instrumentation for testing the weak equivalence principle
  in stratospheric free fall.
  {Rev.\ Sci.\ Instrum.} {\bf 1998}, {\em 69}, 4146.

\bibitem{ffbounce}
  Reasenberg, R.D.; Phillips, J.D.
  A Laboratory Test of the Equivalence Principle as Prolog
  to a Spaceborne Experiment.
  {\em Int.\ J.\ Mod.\ Phys.\ D} {\bf 2007}, {\em 16}, 2245.

\bibitem{ffsr}
  Reasenberg, R.D.; Patla, B.R.; Phillips, J.D.; Thapa, R.
  Design and characteristics of a WEP test in a sounding-rocket payload.
  {\em Class.\ Quant.\ Grav.} {\bf 2012}, {\em 29}, 184013.

\bibitem{planetangles}
  The Euler angles are estimated
  using the data available at
  https://ssd.jpl.nasa.gov/horizons.cgi.

\bibitem{iorio}
  Iorio, L.
  Orbital effects of Lorentz-violating Standard Model Extension
  gravitomagnetism around a static body: a sensitivity analysis.
  {\em Class.\ Quant.\ Grav.} {\bf 2012} {\em 29}, 175007.

\bibitem{hees}
  Hees, A.; Bailey, Q.G.; Le Poncin-Lafitte, C.; Bourgoin, A.; Rivoldini, A.;
  Lamine, B.; Meynadier, F.; Guerlin, C.; Wolf, P.
  Testing Lorentz symmetry with planetary orbital dynamics.
  {\em Phys.\ Rev.\ D} {\bf 2015}, {\em 92}, 064049.


\bibitem{llr1}
  Battat, J.B.R.; Chandler, J.F.; Stubbs, C.W.
  Testing for Lorentz Violation: Constraints on
  Standard-Model Extension Parameters via Lunar Laser Ranging.
  {\em Phys.\ Rev.\ Lett.} {\bf 2007}, {\em 99}, 241103.
\bibitem{llr2}
  Bourgoin, A.; Hees, A.; Bouquillon, S.; Le Poncin-Lafitte, C.;
  Francou, G.; Angonin, M.C.
  Testing Lorentz symmetry with Lunar Laser Ranging.
  {Phys.\ Rev.\ Lett.} {\bf 2016}, {\em 117}, 241301.
  
\bibitem{llr3}
  Bourgoin, A.; Le Poncin-Lafitte, C.; Hees, A.; Bouquillon, S.;
  Francou, G.; Angonin, M.C.
  Lorentz Symmetry Violations from Matter-Gravity Couplings
  with Lunar Laser Ranging.
  {\em Phys.\ Rev.\ Lett.} {\bf 2017}, {\em 119}, 201102.

\bibitem{jennings}
  Jennings, R.J.; Tasson, J.D.; Yang, S.
  Matter-Sector Lorentz Violation in Binary Pulsars.
  {\em Phys.\ Rev.\ D} {\bf 2015}, {\em 92}, 125028.

\bibitem{shao_grav1}
  Shao, L.
  Tests of local Lorentz invariance violation of gravity in
  the standard model extension with pulsars.
  {\em Phys.\ Rev.\ Lett.} {\bf 2014}, {\em 112}, 111103.
\bibitem{shao_grav2}
  Shao, L.
  New pulsar limit on local Lorentz invariance violation
  of gravity in the standard-model extension.
  {\em Phys.\ Rev.\ D} {\bf 2014}, {\em 90}, 122009.
\bibitem{shao_d5}
  Shao, L; Bailey, Q.G.
  Testing velocity-dependent CPT-violating gravitational forces with radio pulsars.
  {\em Phys.\ Rev.\ D} {\bf 2018}, {\em 98}, 084049.
\bibitem{shao_d8}
  Shao, L.; Bailey, Q.G.
  Testing the Gravitational Weak Equivalence Principle
  in the Standard-Model Extension with Binary Pulsars.
  {\em Phys.\ Rev.\ D} {\bf 2019}, {\em 99}, 084017.

\bibitem{pulsar2}
  Shao, L.; Caballero, R.N.; Kramer, M.;
  Wex, N.; Champion, D.J.; Jessner, A.
  A new limit on local Lorentz invariance violation
  of gravity from solitary pulsars.
  {\em Class.\ Quant.\ Grav.} {\bf 2013}, {\em 30}, 165019.

\bibitem{quartz2}
  Goryachev, M.; Kuang, Z.; Ivanov, E.N; Haslinger, P.; Muller, H.; Tobar, M.E.
  Next Generation of Phonon Tests of Lorentz Invariance using Quartz BAW Resonators.
  {\em IEEE Trans.\ on UFFC} {\bf 2018}, {\em 65}, 991.


  
\end{thebibliography}
\end{document}